\documentclass[11pt,twoside,a4paper]{article}
\usepackage[utf8]{inputenc}
\usepackage{latexsym,amsthm}
\usepackage{graphics,physics}
\usepackage{graphicx}
\usepackage{epsfig}
\usepackage{multirow}
\usepackage{amsmath,amssymb,bm}
\usepackage{hhline}
\usepackage{hyperref}
\usepackage{enumitem}
\usepackage{amsfonts}
\usepackage{xcolor}
\usepackage{soul}
\usepackage{bbm}
\usepackage[font=small,labelfont=bf]{caption}
\usepackage[compat=1.1.0]{tikz-feynman}
\usepackage{subcaption}
\usepackage{hhline}
\usepackage{bm,ulem}
\usepackage{subcaption}
\usepackage{float}
\usepackage{feynarts}
\restylefloat{table}
\usepackage[margin=1in]{geometry}
\usepackage[compat=1.1.0]{tikz-feynman}
\usepackage{braket}
\usepackage{nccmath}
\usepackage{cite}
\usepackage{multirow}
\usepackage{makecell}
\def\plusheight{-\the\dimexpr\fontdimen22\textfont2\relax}
\usepackage{amsmath, amsthm}
\usepackage{subcaption}
\usepackage{multirow}
\usepackage{booktabs}
\usepackage{amsmath}
\usepackage{amsfonts}
\usepackage{amssymb}
\usepackage{graphicx}
\usepackage{indentfirst}
\usepackage{slashed}
\usepackage{mathrsfs}
\usepackage{feynmf}
\usepackage{comment}
\usepackage{mathtools}
\usepackage[export]{adjustbox}
\usepackage{tabularray}
\usepackage{bigcenter}

\newcommand\R{{\mathbbm{R}}}

\newcommand{\bea}{\begin{eqnarray}}
	\newcommand{\eea}{\end{eqnarray}}
\newcommand{\bean}{\begin{eqnarray*}}
	\newcommand{\eean}{\end{eqnarray*}}

\def\d{{\rm d}}

\def\Label#1{\label{#1}
	\smash{\hbox to0pt{\raise1ex\hbox{\tiny[#1]}\hss}}}

\def\beq{\begin{equation}}
	\def\eeq{\end{equation}}
\def\bsp#1\esp{\begin{split}#1\end{split}}

\title{\bf Triangular tessellations of one-loop scattering amplitudes in $\phi^3$ theory}
\author{\bf Abhijit B. Das}
\date{%
	Centre for High Energy Physics,\\ Indian Institute of Science,\\ Bangalore-560012, Karnataka, India
	\\[\baselineskip]
	 }

\begin{document}

\maketitle

\begin{abstract}
     Inspired by the recent work of Nima Arkani Hamed and collaborators who introduced the notion of positive geometry to
account for the structure of tree-level scattering amplitudes in bi-adjoint $\phi^3$ theory, which led to
one-loop descriptions of the integrands. Here we consider the one-loop integrals themselves in $\phi^3$
theory. In order to achieve this end, the geometrical construction offered by Schnetz for Feynman diagrams 
is hereby extended, and the results are presented. The extension relies on masking the loop momentum variable with a constant and proceeding with the calculations.
The results appear as a construction given in a diagrammatic manner.   The significance of the resulting triangular diagrams is that they have a common side amongst themselves for the corresponding Feynman diagrams they pertain to. 
Further extensions to this mathematical construction can lead to additional insights into higher loops. A mathematica code has been provided in order to generate the final results given the initial parameters of the theory.

\end{abstract}

\section{Introduction}

Elementary particle physics has been relying on the properties of Feynman diagrams for over half
a century now for precision computations in QED and the Standard Model. They are the fundamental
building blocks for our understanding of interactions. Feynman diagrams are often used to compute scattering
amplitudes in addition to many other observables. Scattering amplitudes are known to satisfy general properties
based on analyticity, unitarity, and crossing symmetry, and there are significant constraints on their behavior in the physical
world. On the other hand, Feynman diagrams also suffer from difficulties of many kinds, including the fact that
they are often divergent, requiring them to be regularized and the theory to be renormalized. This is now a
sophisticated field in itself. Life becomes more complicated as the number of loops increases, as well as the
number of mass scales also increases. Furthermore, Feynman integrals have also opened the gates to research
in the properties of, e.g., hypergeometric functions, and also have enriched subjects as diverse as algebraic geometry,
and their analysis requires knowledge of functional analysis and properties of functions of mathematical physics.
Other approaches to their evaluation analytically in certain limits include, e.g., the Method of Regions, Sector
Decomposition and obtaining so-called Master Integrals and their epsilon expansion have been developed
in differential equations, dispersions relations, etc. In this regard, it must be mentioned that polytopes appear in 
several contexts in the analysis of Feynman integrals \cite{ngon,ads5,ampgeo,polyscat} and for conformal integrals \cite{mellincon,spurious,star,splines}. While finishing our previous works \cite{asymp,hopf}, the prime motivation for this work originated from a paper of Abreu \cite{Abreu} where he considers different kinds of polytopes in his cut analysis.

Scattering amplitudes also have had a new life in the recent past due to the extensive work done on them. Scattering amplitudes in perturbative Quantum Field Theory are obtained by summing up the contributing Feynman diagrams at all orders. Being directly observable, they are entities of prime importance in studying elementary particles in experiments in Large Hadron collider and other particle colliders. An excellent review of recent exciting developments made in the subject is done in  Elvang and Huang \cite{SAIGTAG}, Henn and Plefka \cite{SAIGT}.

However, there are also other rich approaches to their study in terms of the amplituhedron \cite{Amphedron}, associahedron \cite{Arkani-Hamed 2018}. A nice review of these is given in \cite{GGOSA,ampbey}

In this regard, recently Nima Arkani Hamed and his co-workers have advocated the study of scattering amplitudes at tree-level
within the framework of positive geometries. This has led to a very nice geometric picture encapsulated in 
Fig.(16) of their paper \cite{Arkani-Hamed 2018}.  The extension of this to the one-loop integrand has been given in \cite{cluster,Halo1,Halo2} for biadjoint and ordinary $\phi^3$ theory. Also, the extension of these to more general theories has been made in \cite{stokes,accor}

Some years ago, Schnetz \cite{Schnetz(2010)} studied the geometry of one-loop Feynman integrals in $\phi^3$ theory, in which he
introduced the mathematical framework, including hyperbolic geometry. This is based on the earlier work of Davydychev
and Delbourgo, who construct the basic polytopes \cite{Davydychev and Delbourgo(1998)}.

From the above, it may be seen that the issue of one-loop integrals appearing in scattering amplitudes has not
been highlighted much (to be contrasted with the one-loop and higher loop integrands in SUSY and other theories). In this direction, some recent work has been done in \cite{calabi}. The aim of this work is to compute such 
one-loop integrals in a unified framework. And also, subsequently, the prime motivation is to find a geometry for the S-matrix as a whole, uniting all the loop orders, which is possible if we consider the geometry at the integral level.

In order to achieve this end, we construct formalism, which unifies the loop propagators and the internal propagators. What arises
is a generalization of Schnetz that is suitable for representing the scattering amplitude that can be captured in
a diagrammatic form. In the work of Schnetz, he considers only primitive one-loop integrals. In contrast, we are
led to consider all the one-loop diagrams appearing in a scattering amplitude in the theory for the $N=3$ case at $d=3$ dimensions. It is analogous to 
the set of diagrams appearing in the biadjoint $\phi^3$ theory of \cite{Arkani-Hamed 2018}, where only the tree-level amplitudes are considered.

The fundamental difference between our case and the positive geometry case is that here we are working at the level of integrals directly. Also, we are restricting ourselves to the $N=3$ case, where the positive geometry yields a straight line as a special case of the associahedron polytope. It will be interesting to mention that an advancement for finding geometries in the non-perturbative sector has been done in \cite{nonper}.

The scheme of this paper is as follows:
In sec. 2. we start with a short review of Schnetz's work and then proceed to construct our mathematical formalism, which brings together the loop and internal propagators in a single framework. In sec. 3. we evaluate the general quadric surface equation originating from this framework for the case of triangle, bubble, and tadpole diagrams. In sec. 4. we evaluate each diagram and try to give it a geometric representation. In sec. 5. we try to stack up the contributions coming from all the Feynman integrals in order to provide a diagrammatic representation of the scattering amplitude. Sec. 6. contains the discussions and conclusions.

In the Appendix, we explain the equations which are the central figures in the mathematical formalism of this paper. A stand-alone mathematica code is provided, which explicitly codes
the algorithm.

In sec. 2. until eq.(\ref{52}) is a recap of Schnetz. He uses a linear transformation in which the fact that the integration area is a surface integral becomes evident, and the polytope which can be obtained from them becomes obvious. The final result is completely expressed in terms of geometrical quantities in hyperbolic space and the corresponding Euclidean space. Our work extends this work to include other types of one-loop diagrams contributing to the scattering amplitude.

The formalism so far is not enough to address the issue of
the entire set of diagrams appearing in the 1-loop scattering amplitude and the resulting integrals.
Note here that the integrands are considered in \cite{cluster,Halo1,Halo2} where the result appears as polytopes (cluster polytopes and halohedron).
In contrast, when we complete our construction, the results will also appear as polytopes (triangles) of a different
type.

In sec. 3., we apply the actual values of the masking constants $a_i \to 0,1$, to get the results of the corresponding one-loop integrals. Here we discover some new and very interesting facts about the type of hyperbolic geometry obeyed by these integrals. The triangle Feynman diagram for the $N=3$ case consists of all the three involved masking constants $a_i$ having a value equal to 1; for the bubble, any two are equal to 1, and the third one is zero, and for the tadpole, any one masking constant $a_i$ is equal to one and the rest are equal to zero. From this, we obtain a very interesting fact that the triangle Feynman diagram represents the hyperboloid of two sheets, the bubble represents a rotated hyperboloid of two sheets but with one flat axis, and the tadpole represents a rotated hyperboloid of two sheets with both flat axis.

In sec.5.,
 we display the final results. These are displayed by appropriately stacking up the polytopes
that we have obtained. This is the analog of the tree-level bi-adjoint stacking of the polytopes given
by Nima Arkani Hamed. Though not so mathematically rigorous, the fact that the scattering amplitudes have a geometry of their own in this theory is
a significant advance over the simple one-loop formula given by Schnetz.  

Future work could be directed at generalizing this picture to higher loops which could help the computation of scattering amplitudes to higher precision and give a completely new and more straightforward method to evaluate scattering amplitudes involving unique and fascinating geometry. Such a unified framework is in philosophy analogous to that of Nima Arkani Hamed for bi-adjoint
theories, where the tree-level did not require any loop analysis and where the one-loop has been
given only for integrands. In this tree-level context, it has been named the associahedron, whereas, at the one-loop level, the integrand polytope construction is called cluster polytopes. Our work is a
generalization in a different direction up to the first order in the series expansion.

Appendices include the table showing the correspondence of the individual mapped Feynman diagrams with their generating variables, the derivation of certain key equations, and results.

\section{Mathematical Construction}

\subsection{Review}

This is a review of the work done in \cite{Schnetz(2010)}. The one-loop Feynman diagram is given by

\begin{figure}[ht]
\hspace{5cm}
\includegraphics[keepaspectratio=true, height=6cm]{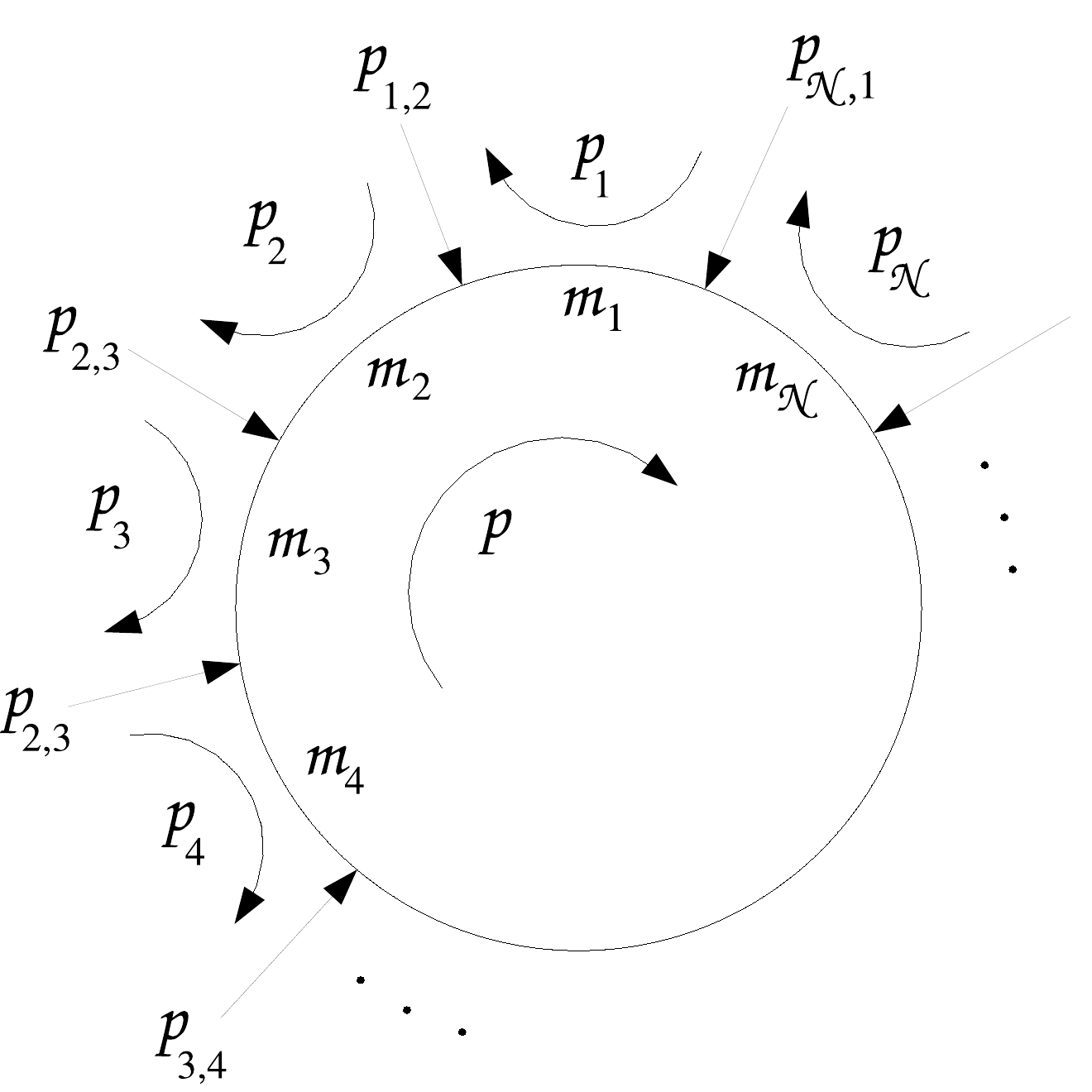}
\caption{Feynman diagram of the one-loop amplitude.}
\end{figure}

\begin{equation}\label{2}
A(p_1,m_1,\dots,p_N,m_N)=\int_{\R^n}{\rm d}^np\frac{1}{Q_1\cdots Q_N}.
\end{equation}

with 

\begin{equation}\label{1}
Q_i=(p-p_i)^2+m_i^2,\quad i=1,\dots ,N.
\end{equation}

where $p$ is the loop momentum. The incoming external momenta are
\begin{equation}\label{1a}
p_{i,i+1}=p_{i+1}-p_i
\end{equation}
where $p_{N+1}=p_1$ and $p_{N,N+1}=p_{N,1}$. Using $\alpha$ parametrization we can write the integrand as
\begin{equation}\label{31}
\frac{1}{Q_1\cdots Q_N}=(N-1)!\int_0^\infty {\rm d}\alpha_2\cdots\int_0^\infty {\rm d}\alpha_N
\frac{1}{(Q_1+\alpha_2Q_2+\ldots+\alpha_NQ_N)^N}.
\end{equation}

Shifting $p\mapsto p+(\sum\alpha_ip_i)/(\sum\alpha_i)$ we get
\begin{equation}\label{34}
\int{\rm d}^np\frac{1}{Q_1\cdots Q_N}=\int_\Delta\Omega\,\frac{2\pi^{\frac{n}{2}}}{\Gamma(\frac{n}{2})}\int_0^\infty\frac{{\rm d}p\,p^{n-1}}
{\left((\sum\alpha_i)p^2+\frac{(\sum\alpha_i)(\sum\alpha_i(p_i^2+m_i^2))-(\sum\alpha_ip_i)^2}{\sum\alpha_i}\right)^N}.
\end{equation}
where
\begin{equation}\label{32}
\Omega=\sum_{i=1}^N(-1)^{i-1}\alpha_i{\rm d}\alpha_1\wedge\ldots\wedge{\rm d}\alpha_{i-1}\wedge{\rm d}\alpha_{i+1}\wedge\ldots\wedge{\rm d}\alpha_N.
\end{equation}
and $\Delta$ is the integration region. The (one-dimensional) $p$-integral can be evaluated and leads to the beta-function with the result
\begin{equation}\label{35}
\int\frac{{\rm d}^np}{Q_1\cdots Q_N}=\pi^{\frac{n}{2}}\Gamma\!\left(N-\frac{n}{2}\right)\!\int_\Delta\!\!\Omega\left(\sum_{i=1}^N\alpha_i\right)^{\!N-n}
\!\left(\sum_{i,j=1}^N\alpha_i\alpha_j\frac{(p_i-p_j)^2+m_i^2+m_j^2}{2}\right)^{\!\frac{n}{2}-N}.
\end{equation}
For the case $N=n$ the term $\left(\sum_{i=1}^N\alpha_i\right)^{\!N-n}$ vanishes and we can just integrate over the surface
\begin{equation}\label{36}
\Delta_1=\left\{\alpha_i:\sum_{i,j=1}^n\alpha_i\alpha_j\frac{(p_i-p_j)^2+m_i^2+m_j^2}{2}=1,\; \alpha_i\geq 0\right\}
\end{equation}
and we obtain
\begin{equation}\label{37}
\int{\rm d}^np\frac{1}{Q_1\cdots Q_n}=\pi^{\frac{n}{2}}\Gamma\left(\frac{n}{2}\right)\int_{\Delta_1}\Omega.
\end{equation}
and finally after change of variables from $\alpha_i$ to $u_i$ as instructed in \cite{Schnetz(2010)} we get 
\begin{equation}\label{52}
\int{\rm d}^np\frac{1}{Q_1\cdots Q_n}=\frac{\hbox{vol}(S^n_{1/2})\hbox{vol}_{H^{n-1}}(\Sigma)}{r\,\hbox{vol}_{\R^{n-1}}(\Sigma)}.
\end{equation}

Note that eq.(\ref{37}) is equivalent to (see for e.g. \cite{Davydychev and Delbourgo(1998)})

\begin{equation}\label{equidelta_1}
    \int{\rm d}^np\frac{1}{Q_1\cdots Q_n}=\pi^{\frac{n}{2}}\Gamma\left(\frac{n}{2}\right)\int_0^{\infty}\cdots\int_0^{\infty}\prod{\rm d}\alpha_i\delta\left(\sum_{i,j=1}^n\alpha_i\alpha_j\frac{(p_i-p_j)^2+m_i^2+m_j^2}{2}-1\right)
\end{equation}

\subsection{The main construction}

\subsubsection{For the $N=3$ case}

Using the result eq.(\ref{52}), we can try to add the Feynman diagrams to find out the overall geometry of the Scattering amplitudes. Here we do the analysis for the $N=3$ case but try to rewrite it as a general $N$ case so that it can be easily extended to the general cases in a future work. In order to do so, we need to sum up the contributions of the tadpoles, the bubbles, and the triangle for the $N=3$ case. The primary challenge in doing so is that using eq.(\ref{52}), the tadpoles and the bubbles will be related to a point and a line, respectively, but the triangle is associated with a triangle itself using this geometrical argument which is a 2-dimensional object in contrast to the one-dimensional bubble and the 0-dimensional tadpole. So we need to construct a mathematical framework such that the bubble and the tadpole can be elevated to the space of a 2-dimensional triangle.
	\begin{figure}[]
		\centering
		\includegraphics[keepaspectratio=true, height=10cm]{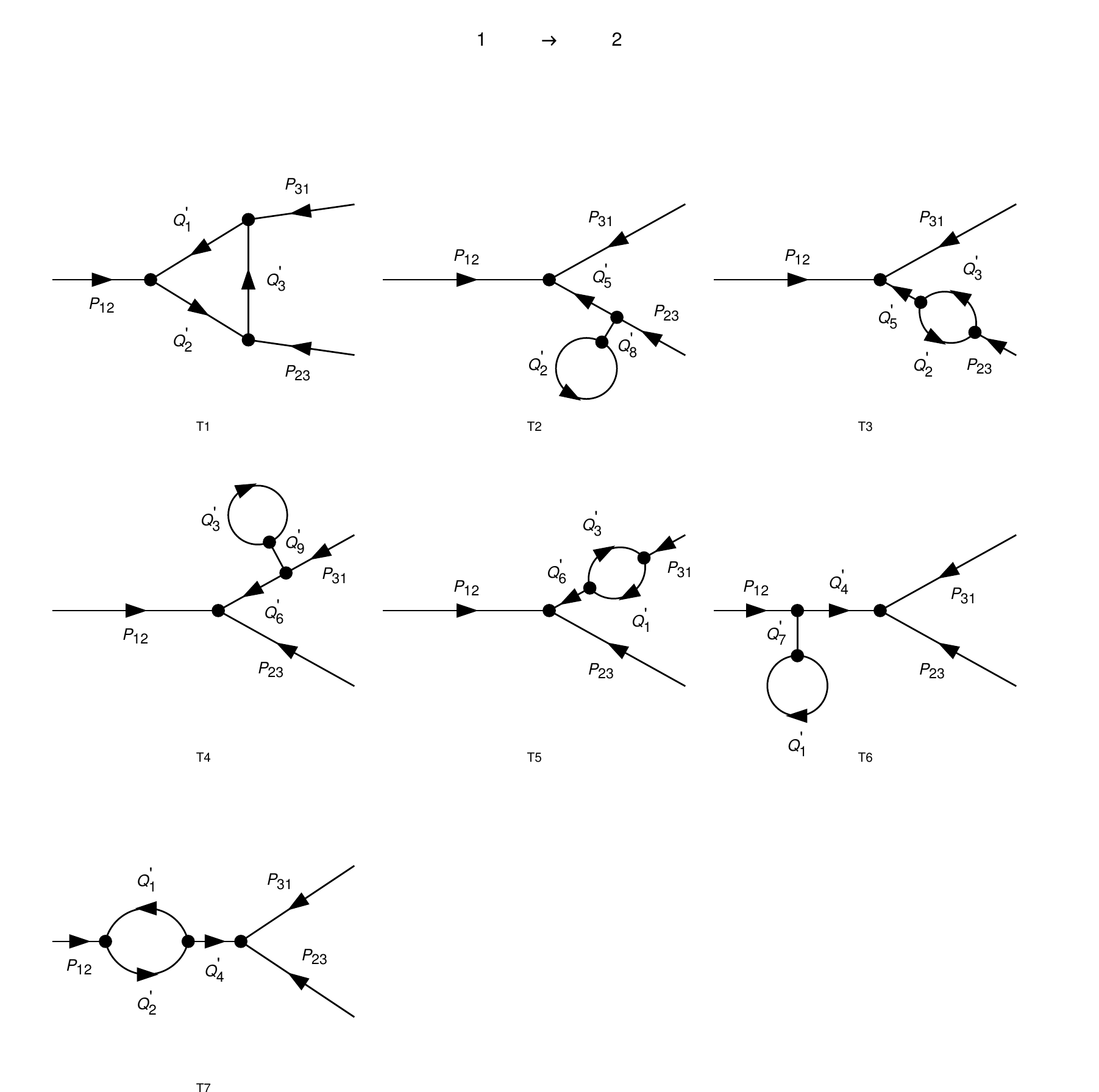}
		\caption{$N=3$ scattering}\label{fig1}
	\end{figure}

For this, we use the property for $N=3$ scattering the sum total of the loop propagators, and the internal propagators is always the same. For example, for the tadpole diagram, we have two internal propagators and one loop propagator (see T2, T4, and T6 in Fig.(\ref{fig1}), the bubble has one internal propagator and two loop propagators (T3, T5, and T7) and the triangle has three loop propagators (T1) all of them having the sum total of 3 loop and internal propagators combined.

We rewrite the loop and internal propagators in a single framework as a part of the loop integral as follows: 

\begin{equation}\label{inloop}
Q_i'=(a_i p-p_i')^2+m_i'^2,\quad i=1,\dots ,3N.
\end{equation}

where

\begin{equation}\label{defpdash}
   p'_i=
\begin{cases}
    p_i,& \quad   i=1,\dots ,N\\
    p_{i-N,i-N+1},   & \quad  i=N+1,\dots ,2N \\
    0 & \quad i=2N+1,\dots,3N
\end{cases}
\end{equation}

and

\begin{equation}
   a_i=
\begin{cases}
    1,& \quad   i=1,\dots ,N\\
    0,   & \quad  i=N+1,\dots ,3N
\end{cases}
\end{equation}

Here we define the sets $I = \{i,i=1,2,..,3N\}$ and $I' \subset I$ such that there exists at least one element $i$ in the range 1 to $N$ and $|I'|=N$. For example $I'=\{i=2,i=7,i=9\}$ is allowed, but $I'=\{i=4,i=5,i=6\}$ or $I'=\{i=1,i=2,i=5,i=6\}$ is not allowed for the $N=3$ case. Thus the range of $I'$ covers all the Feynman integrals contributing to the one-loop scattering amplitude. The labelled $Q_i'$ according to the above formalism is shown in Fig.(\ref{fig1}). Note that the external propagators are not included in this formalism and hence indicated by external momenta. And in place of eq.(\ref{31}), for a particular Feynman diagram in case of $N$ particle scattering we have

\begin{equation}\label{newintegrand}
\prod_{i\in I'}\frac{1}{Q_i'}=(N-1)! \int_0^\infty \prod_{i\in I'} {\rm d}\alpha_i
\frac{\delta(1-\sum_{i\in I'}\alpha_i)}{(\sum_{i\in I'}\alpha_iQ_i')^N}.
\end{equation}

In order to simplify the integral we will do a similar kind of shift as we did earlier. Shifting $p\mapsto p+(\sum\alpha_ia_ip_i)/(\sum\alpha_ia_i^2)$ we get

\begin{equation}\label{shiftedp}
\int{\rm d}^np\prod_{i\in I'}\frac{1}{Q_i'}=\int_\Delta\Omega\,\frac{2\pi^{\frac{n}{2}}}{\Gamma(\frac{n}{2})}\int_0^\infty\frac{{\rm d}p\,p^{n-1}}
{\left((\sum\alpha_ia_i^2)p^2+\frac{(\sum\alpha_ia_i^2)(\sum\alpha_i(p_i'^2+m_i'^2))-(\sum\alpha_ia_ip_i')^2}{\sum\alpha_ia_i^2}\right)^N}.
\end{equation}

After evaluating the $p$-integral we get 

\begin{align}\label{newregion}
& \int{\rm d}^np\prod_{i\in I'}\frac{1}{Q_i'}=  \nonumber \\
& \pi^{\frac{n}{2}}\Gamma\!\left(N-\frac{n}{2}\right)\!\int_\Delta\!\!\Omega\left(\sum_{i=1}^N\alpha_ia_i^2\right)^{\!N-n}
\!\left(\sum_{i,j=1}^N\alpha_i\alpha_j\frac{(a_jp_i'-a_ip_j')^2+a_j^2m_i'^2+a_i^2m_j'^2}{2}\right)^{\!\frac{n}{2}-N}
\end{align}

Again we will consider the case $N=n$ and we are left with integrating over the surface

\begin{equation}\label{Delta'}
\Delta'=\left(\sum_{i,j=1}^N\alpha_i\alpha_j\frac{(a_jp_i'-a_ip_j')^2+a_j^2m_i'^2+a_i^2m_j'^2}{2}=1,\; \alpha_i\geq 0\right\}
\end{equation}

and thus

\begin{equation}\label{beforetrans}
\int{\rm d}^np\prod_{i\in I'}\frac{1}{Q_i'}=\pi^{\frac{n}{2}}\Gamma\left(\frac{n}{2}\right)\int_{\Delta'}\Omega.
\end{equation}

Again this equation is equivalent to 

\begin{equation}\label{equidelta_2}\textstyle
        \int{\rm d}^np\prod_{i\in I'}\frac{1}{Q_i'}=\pi^{\frac{n}{2}}\Gamma\left(\frac{n}{2}\right)\int_0^{\infty}\cdots\int_0^{\infty}\prod{\rm d}\alpha_i\delta\left(\sum_{i,j=1}^N\alpha_i\alpha_j\frac{(a_jp_i'-a_ip_j')^2+a_j^2m_i'^2+a_i^2m_j'^2}{2}-1\right)
\end{equation}

Now, as in \cite{Schnetz(2010)}, in order to write the expression inside eq.(\ref{Delta'}) as a dot product of two vectors, we will use the following equation

\begin{equation}\label{defineconstantr}
a_i^2(m_j'^2+p_j''^2)+ a_j^2(m_i'^2+p_i''^2)= (a_i^2+a_j^2)r_{BT}^2+2a_ia_jr^2
\end{equation}

The definitions of $r_{BT}$ and $r$ alongwith the derivation of this equation is given in the appendix. Here the momentum variables $p_i''$ are given by

\begin{equation}\label{defpddash}
   p_i''^2=
\begin{cases}
    p_i^2,& \quad   i=1,\dots ,N\\
    (p_{i-N,i-N+1}-r_{B})^2,   & \quad  i=N+1,\dots ,2N \\
    r_{BT}^2-m_i'^2 & \quad i=2N+1,\dots,3N
\end{cases}
\end{equation}

where $r_B$ is also defined in the appendix.
Using this we get

$$ \frac{(a_jp_i''-a_ip_j'')^2+a_j^2m_i'^2+a_i^2m_j'^2}{2}  = $$

\begin{align}\label{shiftedr}
 \left[a_jp_i''-ia_j\left(\sqrt{r^2+r^2_{BT}}\right)\right]\cdot \left[a_ip_j''-ia_i\left(\sqrt{r^2+r^2_{BT}}\right)\right]  + \frac{(a_i-a_j)^2r^2_{BT}}{2}
\end{align}

 Here we have chosen $c=0$ in eq.(34) of \cite{Schnetz(2010)} which according to the explanation there is arbitrary. So here we get the momentum variables shifted for the case $a_i=0$ and hence the actual value of the bubble and tadpole propagators also gets shifted to $r^2_{BT}$ !. In order to deal with this we will add the following term to both the sides of eq.(\ref{shiftedr})
 
$$\frac{-(a_i^2+a_j^2)r^2_{BT}+ a_i^2r_j^2+a_j^2r_i^2}{2}
$$

 which gives

 $$ \frac{(a_jp_i''-a_ip_j'')^2+a_j^2m_i'^2+a_i^2m_j'^2-(a_i^2+a_j^2)r^2_{BT}+ a_i^2r_j^2+a_j^2r_i^2 }{2} = $$

 $$
 \frac{(a_jp_i'-a_ip_j')^2+a_j^2m_i'^2+a_i^2m_j'^2}{2}=
 $$

\begin{align}\label{finalshift}
 \left[a_jp_i''-ia_j\left(\sqrt{r^2+r^2_{BT}}\right)\right]\cdot \left[a_ip_j''-ia_i\left(\sqrt{r^2+r^2_{BT}}\right)\right]  - a_ia_jr^2_{BT}+\frac{a_i^2r_j^2+a_j^2r_i^2}{2}
\end{align}
 
 where $r_i$ and $r_j$ are once again defined in appendix. So the LHS of this equation when traced back  gives the original value of the propagators. Thus if we define the new linear transformation here it will be 

\begin{equation}\label{definev}
v^{(k)}=\sum_{i,j=1}^n\alpha_j\left[a_ip_j''-ia_iR\right]^{(k)},\quad k=1,\ldots,n,
\end{equation}

where 

\begin{equation}
    R=\sqrt{r^2+r^2_{BT}}
\end{equation}

 Now if we define a inner product to be of the form (it is also called the Frobenius inner product)

\begin{equation}\label{innerprod}
   \langle v|v \rangle = u'^2 = \sum_{i,j=1}^n\alpha_i\alpha_j \left[a_ip_i''-ia_iR\right]\cdot \left[a_jp_j''-ia_jR\right]
\end{equation}

where

 \begin{equation}\label{defineu}
u'^{(k)}=\sum_{i=1}^n\alpha_i\left[a_ip_i''-ia_iR\right]^{(k)},\quad k=1,\ldots,n,
\end{equation}

is equivalent to the $u$ defined in \cite{Schnetz(2010)}. The significance behind defining two different type of vectors $v$ and $u'$ will be explained in a while. Now if 

\begin{equation}\label{defP'ij}
    P'_{ij} = \left[a_ip_j''-ia_iR)\right]
\end{equation}

we have 

\begin{equation}\label{conic}
   \langle v|v \rangle = u'^2 = \sum_{i,j=1}^n\alpha_i\alpha_j P'_{ii}\cdot P'_{jj} = -1 + \alpha_i\alpha_j\left[a_ia_jr^2_{BT}-\frac{a_i^2r_j^2+a_j^2r_i^2}{2}\right]
\end{equation}

then we get a similar kind of situation as in \cite{Schnetz(2010)} but this time the normal inner product getting converted to Frobenius product if we consider the $v$ variables and along with a additional shift term as seen in the RHS of the above equation. Here eq.(\ref{conic}) represents a general quadric surface equation where the hyperbolic geometry of \cite{Schnetz(2010)} is a special case of this general quadric surface geometry as we will see in the next section. Here we have the spanning Minkowski space vectors given by

\begin{equation}
    v_{j}^{\rm M}=\left(\sum_{i=1}^nP_{ij}^{\rm 'M}\right)/\sum_{i=1}^na_im'_j
\end{equation}

with

\begin{equation}
    P_{ij}^{\rm 'M} = (a_iR,a_ip_j^{''M})
\end{equation}

and

\begin{equation}
    \left(\sum_{i=1}^nP_{ij}^{\rm 'M}\right)^2 = -\sum_{i=1}^na_i^2m_j'^2
\end{equation}

 Thus following similar steps as in \cite{Schnetz(2010)}  finally eq.(\ref{beforetrans}) becomes 

\begin{equation}\label{maineq}
\int{\rm d}^np\prod_{i\in I'}\frac{1}{Q_i'}=\frac{\hbox{vol}(S^n_{1/2})\hbox{vol}_{C^{n-1}}(\Sigma(v_i^{\rm M}))}{R\,\hbox{vol}_{\R^{n-1}}(\Sigma(p_i^{\rm M}))(\sum_ia_i)^n}.
\end{equation}

Here $C^{n-1}$ represents the general quadric surface geometry in $n-1$ dimensions. In the following sections, we will see what type of geometry it leads to for different cases. Note that there is a $\sum_ia_i$ term in the denominator because in the change of variables linear transformation we have chosen the $v_i$ variables which have this factor inside them. If we would have chosen the $u'_i$ variables then it would have been a product of all $a_i$ i.e. $\prod_ia_i$. This is the reason we have to define a different set of variables $v$ other than $u'$. 

As we said, the factor$\sum_ia_i$ is inside and can be taken out common from the $v_i$ variables. So first of all we rewrite eq.(\ref{definev}) as 

\begin{equation}
v^{(k)}=\left(\sum_{i=1}^na_i\right)\sum_{j=1}^n\alpha_j\left[p_j''-iR\right]^{(k)},\quad k=1,\ldots,n,
\end{equation}

We can make the calculations simpler by defining

\begin{equation}\label{definevdash}
v'^{(k)}=\frac{v^{(k)}}{\left(\sum_{i=1}^na_i\right)}=\sum_{j=1}^n\alpha_j\left[p_j''-iR\right]^{(k)},\quad k=1,\ldots,n,
\end{equation}

because of which eq.(\ref{maineq}) takes the form

\begin{equation}\label{maineq_1}
\int{\rm d}^np\prod_{i\in I'}\frac{1}{Q_i'}=\frac{\hbox{vol}(S^n_{1/2})\hbox{vol}_{C^{n-1}}(\Sigma(v_i'^{\rm M}))}{R\,\hbox{vol}_{\R^{n-1}}(\Sigma(p_i^{\rm ''M}))}.
\end{equation}

Thus this mathematical framework has put the bubbles and the tadpoles on the same footing as the triangle and using this equation we can construct an Euclidean triangle and a corresponding quadric triangle corresponding to each diagram in Fig.(\ref{fig1})as follows.

\subsection{Construction of the triangles in mapped space}

We want to see what features the triangles represent when we map the Feynman integrals to them according to this new mathematical formalism. In \cite{Schnetz(2010)}, the triangle diagram T1 in Fig.(\ref{fig1}) is mapped to a triangle using the linear transformation defining the vector $u_i$ but with $p_i$ momenta being the position vectors for the vertices instead of being inside the loop propagator. The original triangle Feynman diagram has no notion of a side length, but here side length and the shape of the mapped triangle are important. Using the formalism presented in this paper, specifically the linear transformation in eq.(\ref{definevdash}), we can map the other types of Feynman diagrams like the bubbles (T3,T5,T7) and the tadpoles (T2,T4,T6) but with different shapes and side lengths.

	\begin{figure}[]
		\centering
		\includegraphics[keepaspectratio=true, height=15cm]{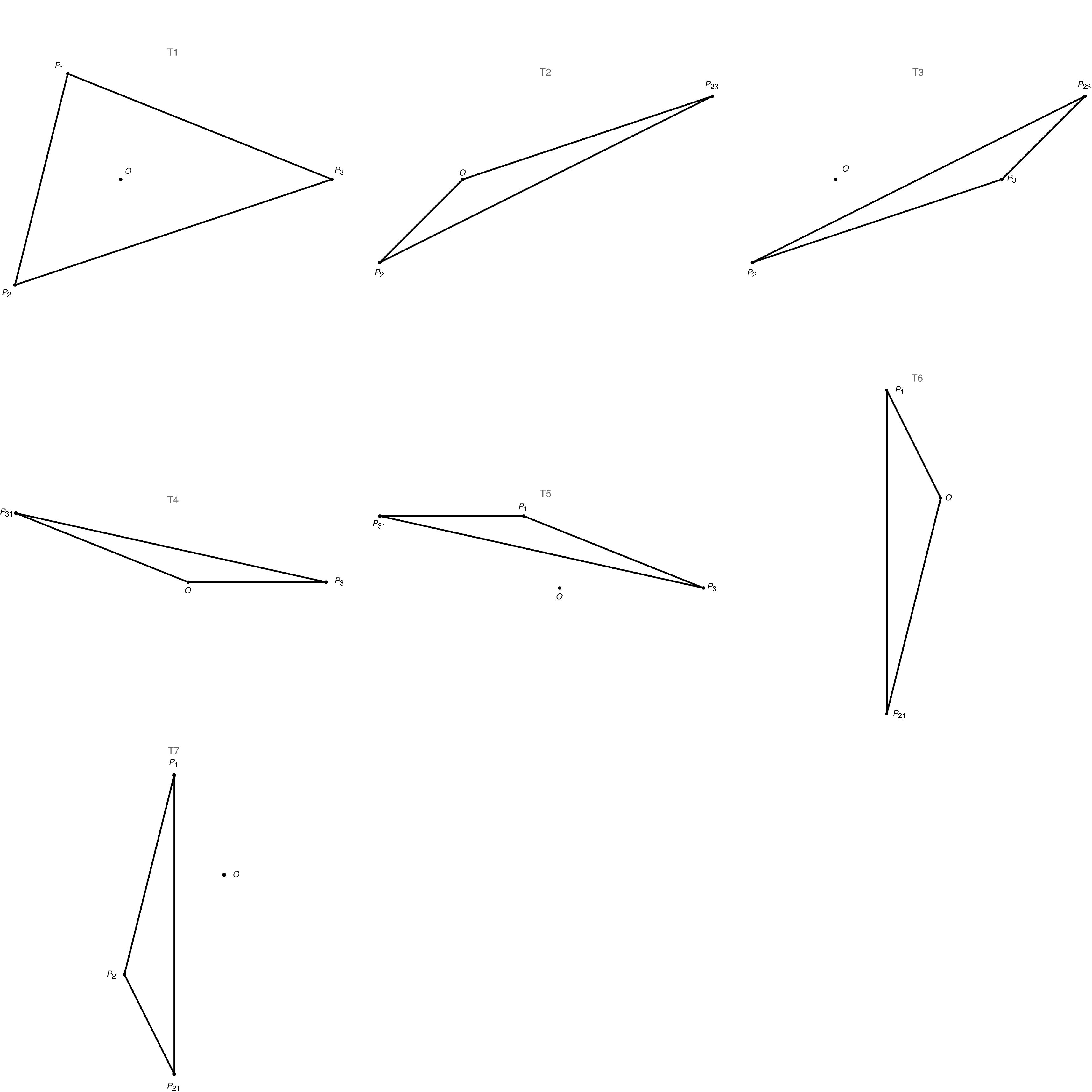}
		\caption{Triangles without shift}\label{fig2}
	\end{figure}

 \subsubsection{Without the shifts}

We will draw the mapped triangles now using eq.(\ref{definevdash}) but first with $p_i'$ defined in eq.(\ref{defpdash}) instead of $p_i''$ defined in eq.(\ref{defpddash}) for brevity and as a first step in understanding. A table, Tab.(\ref{relate}), is provided in the appendix relating the diagrams with the corresponding $Q_i, v_i$ and $p_i$. In the next subsection, we will do it for the actual value $p_i''$. If we denote the incoming momentum in Fig.(\ref{fig1}) as $p_{12}$ and the outgoing momenta as $p_{23}$ and $p_{31}$ then the three vertices of T1 correspond to vectors $p_1$, $p_2$ and $p_3$. For T2, there are two internal propagators with momenta $p_{23}$ and zero,  and one loop propagator with momenta $p_2$ so the corresponding triangles will have vertices $p_{23}$, origin (0) and $p_2$. In this way, we can construct all the triangles in Fig.(\ref{fig2}). Also here we have taken the random example $p_1=\{-1,2\}$, $p_2=\{-2,-2\}$ an $p_3=\{4,0\}$ with $O=\{0,0\}$ as origin for illustration.

So here, we can see that all the triangles have a common side. For example, T1 and T2 have a common side ($p_2-p_3$ and $O-p_{23}$),T2 and T3 have a common side ($p_2-p_{23}$), T1 and T4 have a common side ($p_3-p_1$ and $O-p_{31}$) and so on. This is the main result of our paper to show that using our formalism, we can map the Feynman diagrams to triangles having common sides.

\subsubsection{With shifts}

We can do the same thing as above now with $p_i''$, which gives the original value of propagators. In eq.(\ref{defpddash}), for $i=2N+1,\dots,3N$, the magnitude of $p_i''$ is clear, but the direction is arbitrary. It can be fixed if we follow a particular convention which dictates that $p_i$ and $p_{2N+i}$ become parallel to each other. This specific convention does the precious job of simplifying the calculations and yields the most compact result possible. We have given a detailed discussion about this in Section 5. 

\begin{figure}[]
		\centering
		\includegraphics[keepaspectratio=true, height=15cm]{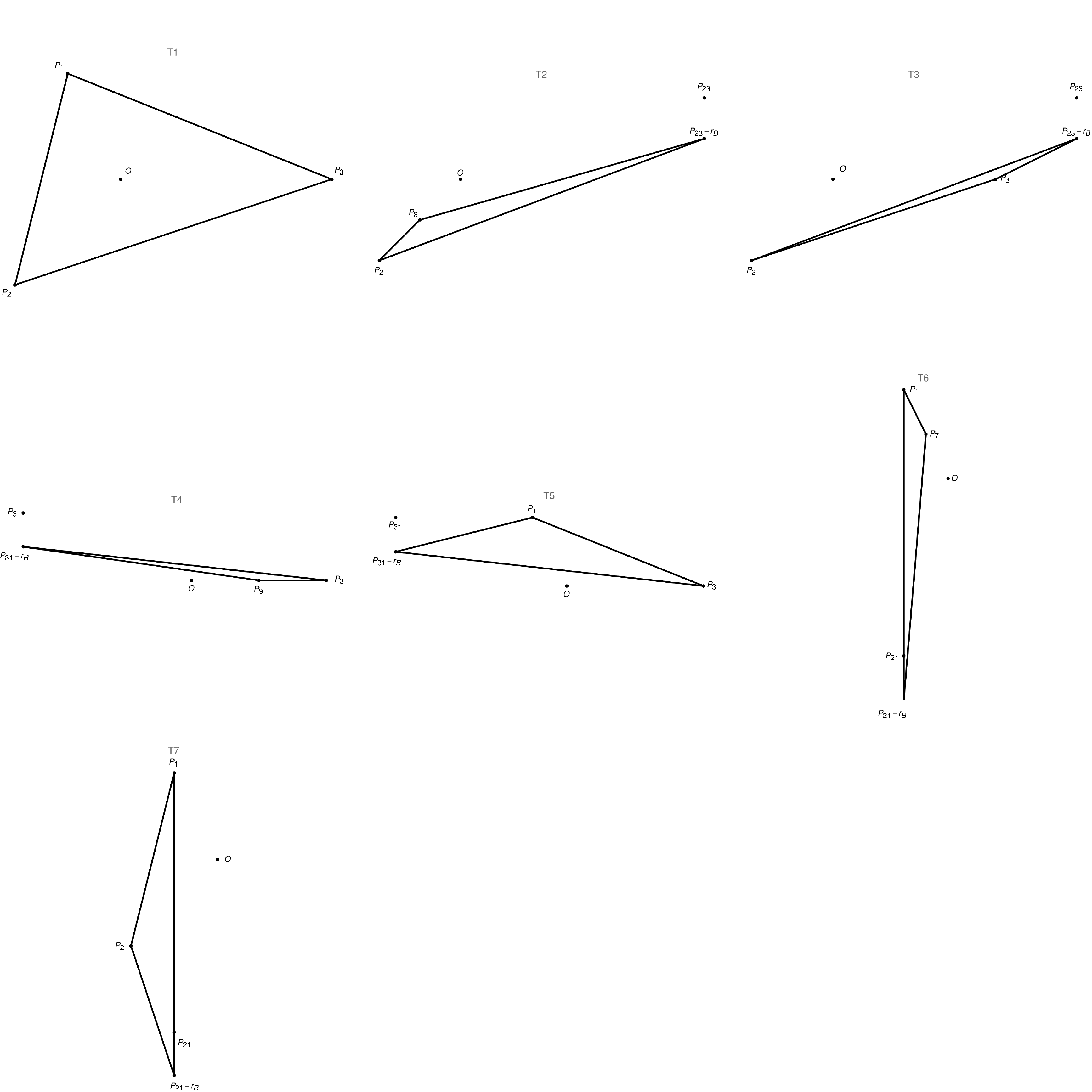}
		\caption{Triangles with shift}\label{fig2shift}
	\end{figure}

After drawing the mapped triangles using eq.(\ref{definevdash}), we see that we still have one side common between triangles, see Fig.(\ref{fig2shift}). For example, T1 and T3 have a common side ($p_2-p_3$), T2 and T3 have a common side ($p_2-p_{23}-r_B$), T1 and T5 have a common side ($p_1-p_3$) and so on. Here we have taken $r_B=\{0,1\}$ and $p_{2N+i}=p_i/2$. So once again, our main argument of showing triangles with common sides is satisfied.

\section{The general quadric surface equation for different cases}

Now in order to evaluate eq.(\ref{maineq_1}) we need to find out the volume of the conic simplex $\Sigma(v_i^{\rm M})$ which is obtained by the intersection of the cone spanned by $\sum_iP'_{ij}$ and the surface governed by eq.(\ref{maineq_1}).

For this we need to find out how to express the general quadric surface equation in terms of the vectors $v'^{(k)}$ in eq.(\ref{definevdash}). This can be done by first expressing the $\alpha_i's$ in terms of these vectors and then substituting in eq.(\ref{innerprod}).

Eq.(\ref{definevdash}) represents a matrix equation and can be solved by taking the inverse \cite{renllc}:

\begin{align}\label{definealpha}
\alpha_j=\sum_{k=1}^n\frac{\varepsilon_{j_1...j_{n-1}j}\varepsilon_{k_1...k_{n-1}k}v'^{(k)}}{(n-1)!\text{det}(p''_1-iR,..,p''_n-iR)}\prod_{l=1}^{n-1}\left[p_{j_l}''-iR\right]^{(k_l)},\quad j=1,\ldots,n, 
\end{align}

where $\varepsilon$ are the generalized Levi-Civita tensors and Einstein summation convention is implied for the $j_l$ and $k_l$ indices. Now using eq.(\ref{innerprod}) and eq.(\ref{definevdash}), eq.(\ref{conic}) can be rewritten in a matrix form as 

\begin{align}\label{defmateqn}\scriptsize
\begin{bmatrix}
\alpha_1[p''_1-iR]& \alpha_2[p''_2-iR]& \cdots & \alpha_n[p''_n-iR]
\end{bmatrix}
\begin{bmatrix}
a_1^2 & a_1 a_2 & \cdots & a_1 a_n \\
a_1 a_2 & a_2^2 & \cdots & a_2 a_n \\
\vdots & \vdots & \ddots & \vdots \\
a_1 a_n & a_2 a_n & \cdots & a_n^2
\end{bmatrix}
\begin{bmatrix}
\alpha_1[p''_1-iR]\\ \alpha_2[p''_2-iR] \\ \vdots \\ \alpha_n[p''_n-iR]
\end{bmatrix}
= \nonumber \\ \scriptsize
-1+
\begin{bmatrix}
\alpha_1 & \alpha_2 & \cdots & \alpha_n
\end{bmatrix}
\left\{
\begin{bmatrix}
a_1^2 & a_1 a_2 & \cdots & a_1 a_n \\
a_1 a_2 & a_2^2 & \cdots & a_2 a_n \\
\vdots & \vdots & \ddots & \vdots \\
a_1 a_n & a_2 a_n & \cdots & a_n^2
\end{bmatrix} -
\begin{bmatrix}
a_1^2 r_1^2 & a_1^2 r_2^2 & \cdots & a_1^2 r_n^2 \\
a_2^2 r_1^2 & a_2^2 r_2^2 & \cdots & a_2^2 r_n^2 \\
\vdots & \vdots & \ddots & \vdots \\
a_n^2 r_1^2 & a_n^2 r_2^2 & \cdots & a_n^2 r_n^2
\end{bmatrix} 
\right\}
\begin{bmatrix}
\alpha_1\\ \alpha_2 \\ \vdots \\ \alpha_n
\end{bmatrix}
\end{align}

where the $\alpha_i's$ are given by eq.(\ref{definealpha}). Note that the last term inside the square bracket in eq.(\ref{conic}) is actually symmetric and hence only one of its parts is written here by dropping the half factor. This is the general quadric surface equation which can be plotted in terms of $v'^{(k)}$. And for finding out the the principal axes of the general quadric surface we need to diagonalize the set of middle matrices in LHS of the above equation with components of $v'$ as  elements in the row and coloumn matrix at the right and left sides of the set of middle matrices respectively.

\begin{align}\label{diamat}
& \hspace{6.7cm} \text{Diagonalize this} \nonumber \\
&
\resizebox{0.85\hsize}{!}{
$
\begin{bmatrix}
v'^{(1)} & v'^{(2)} &
\cdots & 
v'^{(n)}
\end{bmatrix}
\overbrace{\left\{
\begin{bmatrix}
\alpha_1'^{(1)}[p''_1-iR]& \alpha_2'^{(1)}[p''_2-iR]& \cdots & \alpha_n'^{(1)}[p''_n-iR] \\
\alpha_1'^{(2)}[p''_1-iR]& \alpha_2'^{(2)}[p''_2-iR]& \cdots & \alpha_n'^{(2)}[p''_n-iR] \\
\vdots & \vdots & \ddots & \vdots
\\
\alpha_1'^{(n)}[p''_1-iR]& \alpha_2'^{(n)}[p''_2-iR]& \cdots & \alpha_n'^{(n)}[p''_n-iR]
\end{bmatrix}
\begin{bmatrix}
a_1^2 & a_1 a_2 & \cdots & a_1 a_n \\
a_1 a_2 & a_2^2 & \cdots & a_2 a_n \\
\vdots & \vdots & \ddots & \vdots \\
a_1 a_n & a_2 a_n & \cdots & a_n^2
\end{bmatrix}
\begin{bmatrix}
\alpha_1'^{(1)}[p''_1-iR]& \alpha_2'^{(1)}[p''_2-iR]& \cdots & \alpha_n'^{(1)}[p''_n-iR] \\
\alpha_1'^{(2)}[p''_1-iR]& \alpha_2'^{(2)}[p''_2-iR]& \cdots & \alpha_n'^{(2)}[p''_n-iR] \\
\vdots & \vdots & \ddots & \vdots
\\
\alpha_1'^{(n)}[p''_1-iR]& \alpha_2'^{(n)}[p''_2-iR]& \cdots & \alpha_n'^{(n)}[p''_n-iR]
\end{bmatrix}^T \right.}
$} \nonumber \\ & \hspace{2cm}
\resizebox{0.8\hsize}{!}{
$\underbrace{
-\left.\begin{bmatrix}
\alpha_1'^{(1)}& \alpha_2'^{(1)}& \cdots & \alpha_n'^{(1)} \\
\alpha_1'^{(2)}& \alpha_2'^{(2)}& \cdots & \alpha_n'^{(2)} \\
\vdots & \vdots & \ddots & \vdots
\\
\alpha_1'^{(n)}& \alpha_2'^{(n)}& \cdots & \alpha_n'^{(n)}
\end{bmatrix}
\left\{
\begin{bmatrix}
a_1^2 & a_1 a_2 & \cdots & a_1 a_n \\
a_1 a_2 & a_2^2 & \cdots & a_2 a_n \\
\vdots & \vdots & \ddots & \vdots \\
a_1 a_n & a_2 a_n & \cdots & a_n^2
\end{bmatrix} -
\begin{bmatrix}
a_1^2 r_1^2 & a_1^2 r_2^2 & \cdots & a_1^2 r_n^2 \\
a_2^2 r_1^2 & a_2^2 r_2^2 & \cdots & a_2^2 r_n^2 \\
\vdots & \vdots & \ddots & \vdots \\
a_n^2 r_1^2 & a_n^2 r_2^2 & \cdots & a_n^2 r_n^2
\end{bmatrix} 
\right\}
\begin{bmatrix}
\alpha_1'^{(1)}& \alpha_2'^{(1)}& \cdots & \alpha_n'^{(1)} \\
\alpha_1'^{(2)}& \alpha_2'^{(2)}& \cdots & \alpha_n'^{(2)} \\
\vdots & \vdots & \ddots & \vdots
\\
\alpha_1'^{(n)}& \alpha_2'^{(n)}& \cdots & \alpha_n'^{(n)}
\end{bmatrix}^T \right\}}
\begin{bmatrix}
v'^{(1)} \\ v'^{(2)} \\
\vdots \\ 
v'^{(n)}
\end{bmatrix}
=-1
$}
\end{align}

where 

\begin{equation}\label{definealphadash}
\alpha_j'^{(k)}=\frac{\varepsilon_{j_1...j_{n-1}j}\varepsilon_{k_1...k_{n-1}k}}{(n-1)!\text{det}(p''_1-iR,..,p''_n-iR)}\prod_{l=1}^{n-1}\left[p_{j_l}''-iR\right]^{(k_l)},\quad j=1,\ldots,n,
\end{equation}

The diagonalization can be performed by using spectral decomposition where the diagonalizable matrix $A$ can be written as a product of unitary matrix $U$, diagonal matrix $D$, and the transpose of the unitary matrix

\begin{equation}
    A=UDU^T
\end{equation}

where $D$ is the diagonal matrix comprising of the eigenvalues of $A$ and $U$ is the orthonormal eigenvector basis with corresponding eigenvectors as columns of $U$. So if the eigenvalues are given by $\lambda_i$ and the column vectors of $U$ are given by $u_i$ then the general quadric surface becomes \cite{Johnston21}

\begin{equation}\label{eigconic}
    \lambda_1(u_1.v')^2 + \lambda_2(u_2.v')^2 + \cdots + \lambda_n(u_n.v')^2 = -1
\end{equation}

Thus the principal axes of the general quadric surface lie in the direction of $u_i$, and if we choose one of the directions to have the complex part of the momenta ($iR$), for example, if we choose $u_1$, then using eq.(\ref{eigconic}) we can see that the general quadric surface actually satisfies hyperbolic geometry with unequal eigenvalues. 

This is different from the situation in \cite{Schnetz(2010)} in the way that the principal axes of the hyperbolic geometry there was parallel to coordinate axes, and the eigenvalues were of equal magnitude unity. Here the principal axes with different magnitudes of eigenvalues are rotated with respect to the coordinate axes in the case of one or more $a_i=0$.

Now here we have three cases of the general quadric surface, eq.(\ref{eigconic}), as per the type of corresponding Feynman diagram - the triangle, the bubbles, and the tadpoles.

\subsection{Triangle}

For the triangle case we have all $a_i's=1$ and hence using eq.(\ref{definevdash}) and eq.(\ref{defmateqn}) the general quadric surface equation becomes

\begin{equation}
    v'^2 = -1
\end{equation}

For $n=3$, if we choose the complex part of momenta ($iR$) to be in the $v'^{(3)}$ direction, then this denotes the equation of hyperboloid of two sheets

\begin{equation}\label{hyptwosheet}
    (v'^{(1)})^2+(v'^{(2)})^2-(v'^{(3)})^2=-1
\end{equation}

\begin{figure}[ht]
\hspace{3cm}\epsfig{file=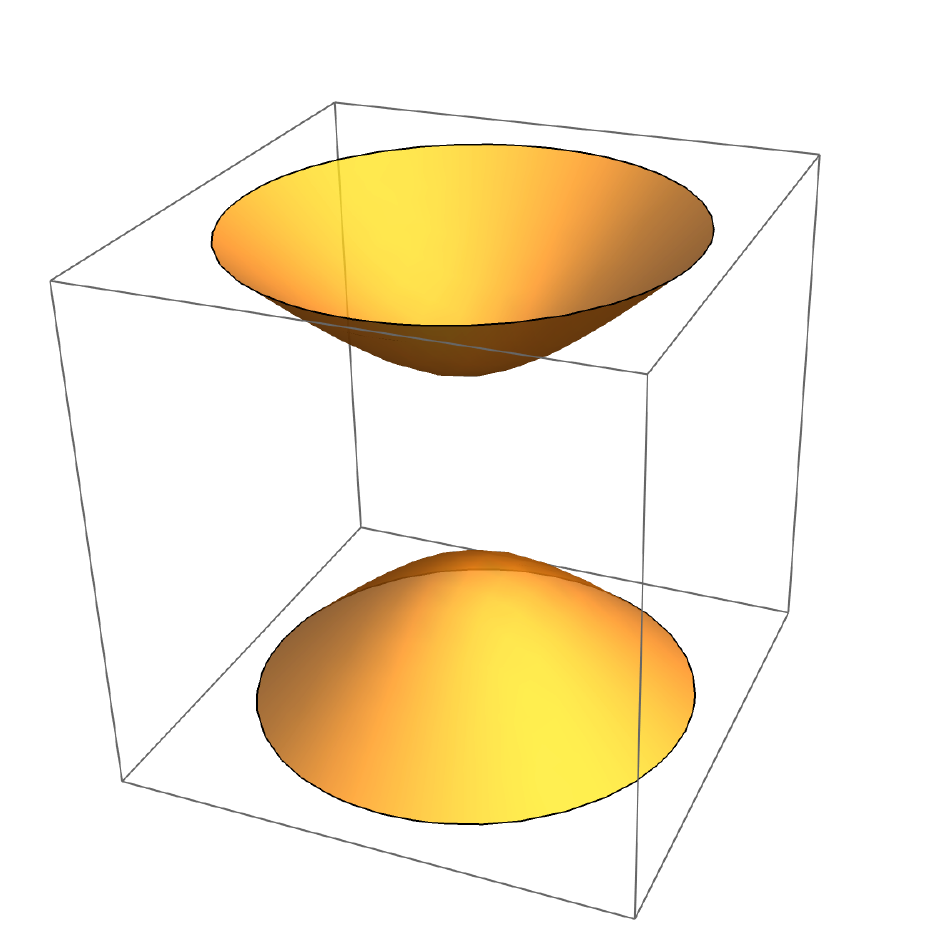,width=10cm}
\caption{ Hyperboloid of two sheets.}\label{hyptwosheetfig}
\end{figure}

So this is a special case where the eigenvalues are equal in magnitude in the general quadric surface equation eq.(\ref{eigconic}). Now according to eq.(\ref{maineq}), we have to find out the volume of the intersection region of the infinite open simplex made by $p_i'$ vectors and the above surface.

\subsection{Bubble}

We get a bubble diagram when any two out of three $a_i's=1$ and the third one equals zero. So in eq.(\ref{defmateqn}), the middle matrix is block diagonal for this case, with any one diagonal entry equal to zero. Because of this, it is not very hard to infer that the eigenvalue matrix $D$ will also be block diagonal in this case, with the corresponding eigenvalue equal to zero.

Using this eq.(\ref{eigconic}) roughly becomes of the form

\begin{equation}\label{bubcon}
    -\lambda_1(u_1.v')^2 + \lambda_2(u_2.v')^2 + \cdots + \lambda_{n-1}(u_{n-1}.v')^2 +  0.(u_n.v')^2 = -1
\end{equation}

\begin{figure}[ht]
\hspace{3cm}\epsfig{file=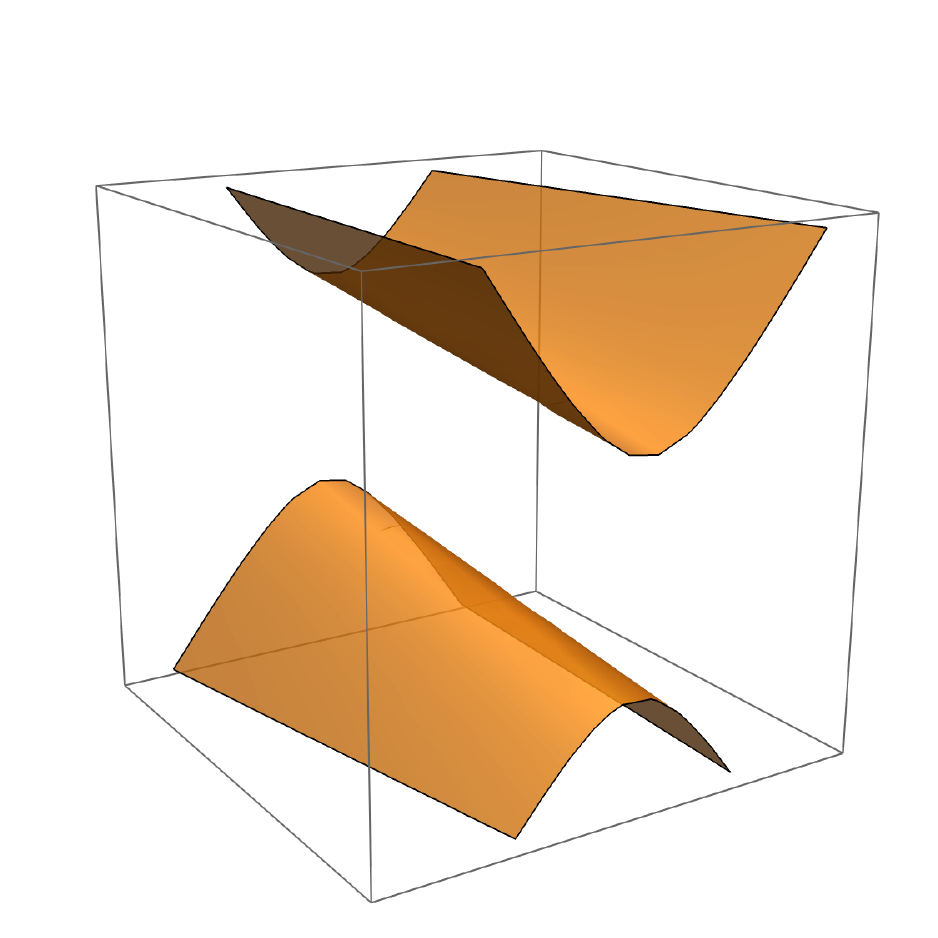,width=10cm}
\caption{ Hyperboloid of two sheets with one flat axis.}\label{hyptwosheetoneflat}
\end{figure}

where the complex part of momenta ($iR$) is chosen along $u_1$ direction, and the nth eigenvalue is chosen to be zero. Here we can see that this equation for $n=3$ pertains to the equation of elliptic hyperboloid of two sheets but with one of the axes of the ellipse having length equal to infinity ($1/\sqrt{\lambda_n}=\infty$). So this looks like a hyperbolic cylinder.

\subsection{Tadpole}

Now for the tadpole case, any two out of three $a_i's$ are equal to zero, and the third one is equal to one. Due to this, once again, the middle matrix in eq.(\ref{defmateqn}) is block diagonal with any two diagonal entries equal to zero. Again due to this, the eigenvalue matrix $D$ becomes block diagonal with any two corresponding eigenvalues equal to zero.

\begin{equation}\label{tadcon}
    -\lambda_1(u_1.v')^2 + \lambda_2(u_2.v')^2 + \cdots + \lambda_{n-2}(u_{n-2}.v')^2 + 0.(u_{n-1}.v')^2 +  0.(u_n.v')^2 = -1
\end{equation}

\begin{figure}[ht]
\hspace{3cm}\epsfig{file=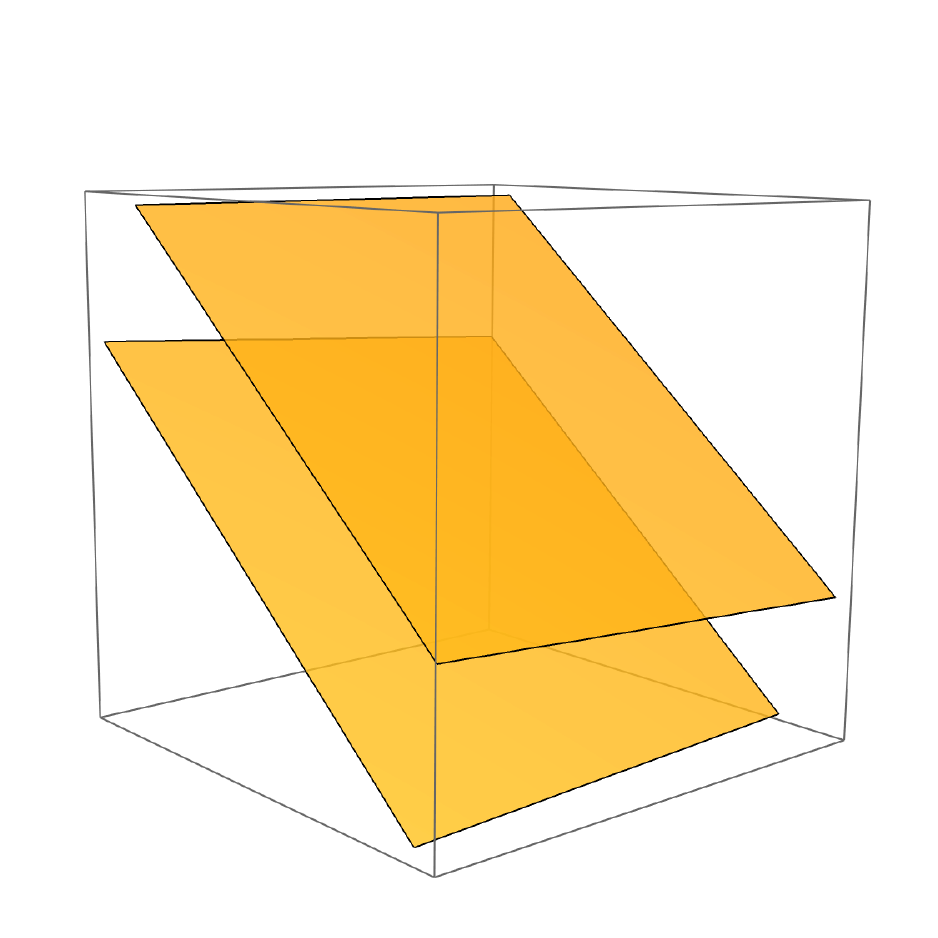,width=10cm}
\caption{ Hyperboloid of two sheets with both flat axis.}\label{twoflat}
\end{figure}

So for the $n=3$ case, this is still the equation for the elliptic hyperboloid but with both the ellipse axes having length equal to infinity. So the cup of the hyperboloid completely flattens down, and what we get are just two completely flat planes.

\section{Evaluation of the Feynman diagrams}

Now the Feynman diagram can be evaluated from the obtained quadrics by finding out the volume of intersection of the quadric with corresponding simplex made by the $p_i''-iR$ vectors. 

For this we need to first find out the volume element ${vol}_{C^{n-1}}(\Sigma(v_i'^{\rm M}))$ in eq.(\ref{maineq_1}). Using eq.(\ref{eigconic}) for the $n=3$ case, it is given by 
\begin{equation}
    {vol}_{C^{n-1}}(\Sigma(v_i'^{\rm M}))=\int \delta \left(    \lambda_1(u_1.v')^2 + \lambda_2(u_2.v')^2 + \lambda_3(u_3.v')^2 +1 \right) dv'^{(1)}dv'^{(2)}dv'^{(3)}
\end{equation}

If we do the $v'^{(3)}$ integration first and if $u_{j}^{(i)}$ is the component of $u_j$ along the direction of $v'^{(i)}$ then we get

{\scriptsize

\begin{align}\label{genvolele}
    & {vol}_{C^{n-1}}(\Sigma(v_i'^{\rm M}))=\int  dv'^{(1)}dv'^{(2)} \times \nonumber \\
    &  \left( \left(\sum_i \lambda_{i}\left(u_i^{(3)}\right)^2\right)\left(\sum_{i}\lambda_{i}\left(\sum_j{u_{i}^{(j)}v'^{(j)}}\right)^2 - \left(\sum_i\lambda_iu_i^{(3)}\left(\sum_j{u_{i}^{(j)}v'^{(j)}}\right)\right)^2\left(\sum_i \lambda_{i}\left(u_i^{(3)}\right)\right)^{-1} +1\right) \right)^{-1}
\end{align}

}%

Now this can be solved using polar co-ordinates

\begin{align}
    v'^{(1)} = \rho \cos{\theta} \\
    v'^{(2)} = \rho \sin{\theta}
\end{align}

where $\rho$ is the radial distance from the origin and $\theta$ is the usual angle in polar co-ordinates. Using this eq.(\ref{genvolele}) becomes 

\begin{equation}
    {vol}_{C^{n-1}}(\Sigma(v_i'^{\rm M}))=\int\frac{\rho\hspace{0.1cm} d\rho\hspace{0.1cm} d\theta}{\sqrt{A[u_i](\rho^2C[\theta]+1)}}
\end{equation}

where

\begin{align}
    &A[u_i]=\left(\sum_i \lambda_{i}\left(u_i^{(3)}\right)^2\right) \\
    \rho^2C[\theta]=\sum_{i}\lambda_{i}\left(\sum_j{u_{i}^{(j)}v'^{(j)}}\right)^2 - & \left(\sum_i\lambda_iu_i^{(3)}\left(\sum_j{u_{i}^{(j)}v'^{(j)}}\right)\right)^2\left(\sum_i \lambda_{i}\left(u_i^{(3)}\right)\right)^{-1}
\end{align}

Now if we define 

\begin{equation}
    \gamma = \frac{\rho}{v'^{(3)}} = \rho\left(\rho B[\theta]+\sqrt{-\frac{\rho^2C[\theta]+1}{A[u_i]}}\right)^{-1}
\end{equation}

with

\begin{equation}
    \rho B[\theta] = - \left(\sum_i\lambda_iu_i^{(3)}\left(\sum_j{u_{i}^{(j)}v'^{(j)}}\right)\right)
\end{equation}

then the integral becomes

\begin{equation}\label{generalintegral}
     {vol}_{C^{n-1}}(\Sigma(v_i'^{\rm M}))=\int\frac{\sqrt{A[u_i]} \gamma  \left(\gamma B[\theta]-1\right) }{\left(A[u_i] \left(\gamma B[\theta]-1\right){}^2+\gamma ^2 C[\theta]\right){}^2}\sqrt{-\frac{\gamma ^2 C[\theta]}{A[u_i] \left(\gamma B[\theta]-1\right){}^2}-1}\hspace{0.1cm} d\gamma d\theta
\end{equation}

The geometrical significance of the variable $\gamma$ will be described in the following subsections. The $\gamma$ integration can be made in a straightforward manner and the result is given by

\begin{equation}
    {vol}_{C^{n-1}}(\Sigma(v_i'^{\rm M}))=\int-\frac{\frac{1}{\sqrt{-\frac{a^2 b^2 C[\theta]}{A[u_i] \left(\csc (\text{$\theta $2}) (b \sin (\theta -\text{$\theta $1}-\text{$\theta $2})-a \sin (\theta -\text{$\theta $1}))+a b B[\theta]\right){}^2}-1}}+i}{\sqrt{A[u_i]} C[\theta]}d\theta
\end{equation}

The limits of this integration have been discussed in detail in the following subsection. The $\theta$ integration is difficult, and we can get a solution in terms of infinite sums of known functions. This is not actually required for the time being, and hence we will just take a series expansion of eq.(\ref{generalintegral}) and then do the integration. So if we do this, we get

\begin{align}\label{seriesgengamma}
       & {vol}_{C^{n-1}}(\Sigma(v_i'^{\rm M})) =  \int\d\gamma d\theta\left\{-\frac{i \gamma }{(A[u_i])^{3/2}}-\frac{3 i \gamma ^2 B[\theta]}{(A[u_i])^{3/2}}-\frac{5 i \gamma ^4 \left(4 A[u_i] b_{\theta }^3-3 B[\theta] C[\theta]\right)}{2 (A[u_i])^{5/2}} \right. \nonumber \\
     &  \left.-\frac{3 i \gamma ^3 \left(4 A[u_i] B[\theta]^2-C[\theta]\right)}{2 (A[u_i])^{5/2}}-\frac{15 i \gamma ^5 \left(8 (A[u_i])^2 B[\theta]^4-12 A[u_i] B[\theta]^2 C[\theta]+C[\theta]^2\right)}{8 (A[u_i])^{7/2}}+\cdots\right.
\end{align}

Now this can be solved order by order in $\gamma$. We have discussed the evaluation for different cases in the following subsections.

\subsection{Triangle}

For the triangle, we inferred that the quadric is actually a hyperboloid of two sheets with principal axis parallel to the axis containing the complex part of the momenta. Hence using eq.(\ref{hyptwosheet}) the ${vol}_{C^{n-1}}(\Sigma(v_i'^{\rm M}))$ in eq.(\ref{maineq_1}) takes the form

\begin{equation}
    {vol}_{C^{n-1}}(\Sigma(v_i'^{\rm M}))=\int \delta \left( (v'^{(1)})^2+(v'^{(2)})^2-(v'^{(3)})^2 + 1 \right) dv'^{(1)}dv'^{(2)}dv'^{(3)}
\end{equation}

which gets simplified to the expression of standard hyperbolic area

\begin{equation}\label{hyptwoarea}
    \int \delta \left( (v'^{(1)})^2+(v'^{(2)})^2-(v'^{(3)})^2 + 1 \right) dv'^{(1)}dv'^{(2)}dv'^{(3)} = \int \frac{dv'^{(1)}dv'^{(2)}}{\sqrt{(v'^{(1)})^2+(v'^{(2)})^2 + 1}}
\end{equation}

The limits of the integration for the intersection region can be simplified if we look at things using the Beltrami-Klein model of hyperbolic geometry. Here we have the Klein disc which maps the entire hyperbolic manifold onto a disk of radius one and at a unit distance height from the origin.

\begin{figure}[ht]
\hspace{3cm}\epsfig{file=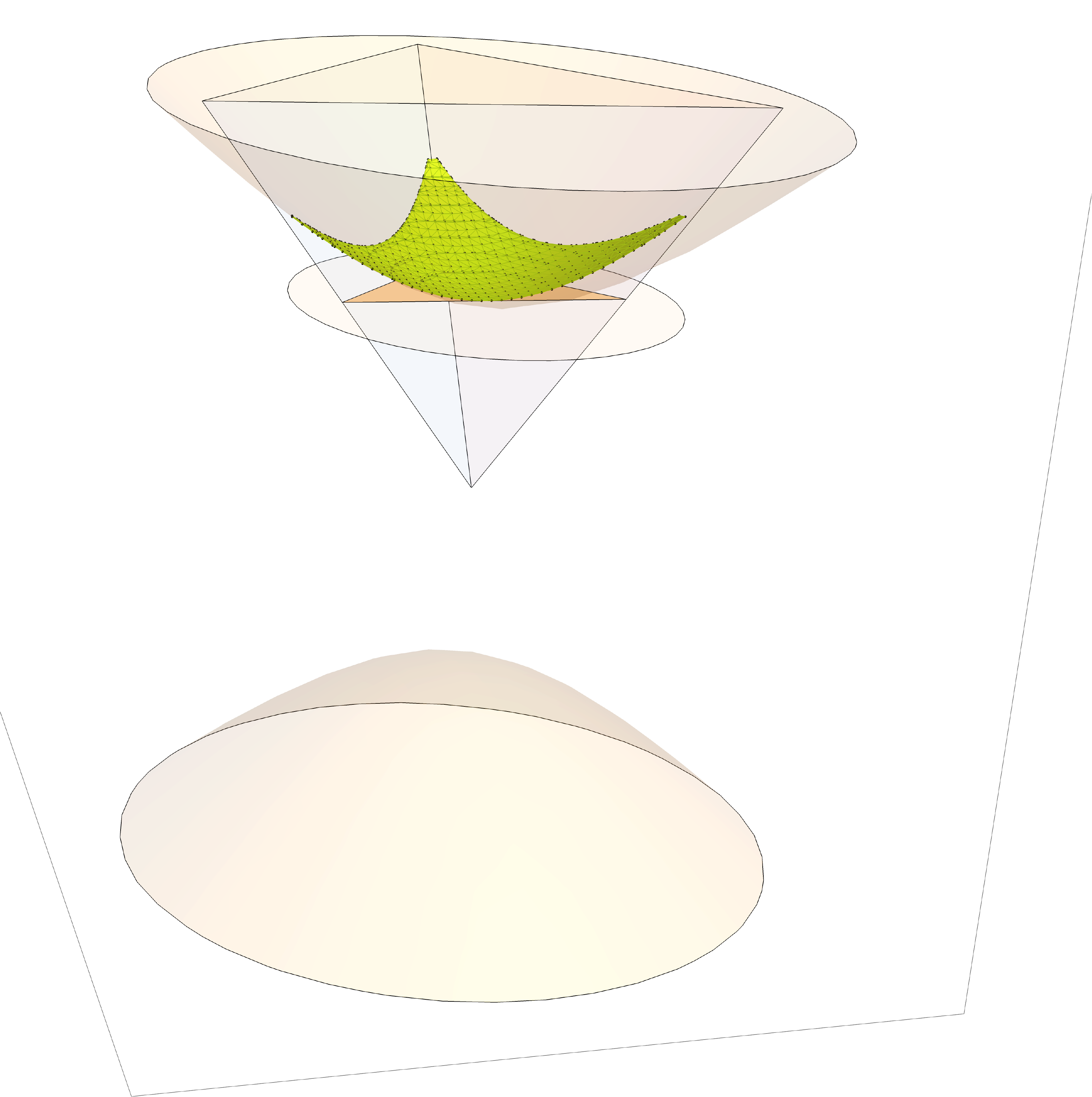,width=10cm}
\caption{Figure showing the intersection of the hyperboloid with the simplex (the green surface) and the intersection of the Klein disc and the simplex (the orange surface)}\label{klein}
\end{figure}

Now if we define the radial distance from the origin

\begin{equation}
    \rho^2 = (v'^{(1)})^2+(v'^{(2)})^2
\end{equation}

and the radial distance of the projection on the Klein disc

\begin{equation}
    \gamma = \frac{\rho}{\sqrt{\rho^2+1}}
\end{equation}

then in terms of these variables, eq.(\ref{hyptwoarea}) becomes

\begin{equation}
    \int \frac{dv'^{(1)}dv'^{(2)}}{\sqrt{(v'^{(1)})^2+(v'^{(2)})^2 + 1}} = \int \frac{\gamma}{\sqrt{(1-\gamma^2)^{3}}}\hspace{0.1cm} d\gamma d\theta
\end{equation}

where $\theta$ is the usual angle in polar co-ordinates.

Now the intersection region or the hyperbolic triangle has a projection which is a Euclidean triangle on the Klein disc. So the limits of integration here are straight lines in the $\gamma$ and $\theta$ coordinate system. So for the triangle diagram, we had constructed a triangle in Fig.(\ref{fig2}), the projection on the Klein disc will also be a  similar triangle with proportional side lengths. 

Now we can split that projected similar triangle into three triangles with common sides having lengths `a', `b', and `c' proportional to the length of momenta $p_1$, $p_2$, and $p_3$, respectively. Also the triangles subtend the angles `$\theta_1$', `$\theta_2$' and `$\theta_3$' on the origin.

Now we will calculate the hyperbolic area of the triangle with side lengths `a' and `b' and angle $`\theta_1$' in between them. For this, the limits of integration for $\gamma$ is between zero to the straight line connecting the tip of the triangle sides `a' and `b', and for $\theta$, it is between $\theta$ to $\theta + \theta_1$.

\begin{equation}
    \int_{\theta}^{\theta + \theta_1} \int_{0}^{\frac{a b \sin (\text{$\theta $2})}{a \sin (\theta -\text{$\theta $1})-b \sin (\theta -\text{$\theta $1}-\text{$\theta $2})}} \frac{\gamma}{\sqrt{(1-\gamma^2)^{3}}}\hspace{0.1cm} d\gamma d\theta
\end{equation}

This matches eq.(\ref{generalintegral}) with $A[u_i]=-1$,$B[\theta]=0$ and $C[\theta]=1$ and it can be evaluated to the expression containing the inverse tan and inverse cot functions

\begin{equation}\scriptstyle
    -\tan ^{-1}\left(\frac{\sqrt{1-a^2} b \sin (\text{$\theta $2})}{a-b \cos (\text{$\theta $2})}\right)-\tan ^{-1}\left(\frac{a \sqrt{1-b^2} \sin (\text{$\theta $2})}{b-a \cos (\text{$\theta $2})}\right)-\cot ^{-1}\left(\frac{a \sin (\text{$\theta $2})}{b-a \cos (\text{$\theta $2})}\right)-\cot ^{-1}\left(\frac{b \sin (\text{$\theta $2})}{a-b \cos (\text{$\theta $2})}\right)
\end{equation}

which gives the standard result of $\pi - \theta_2$ for the area of the `infinite hyperbolic triangle' with two ideal vertices in the limit $a , b \to 1$.

This result can be expressed in terms of the Euclidean triangle area $a b sin(\theta_2)$, if we rewrite hyperbolic area inetgral as 

\begin{equation}
    F[k] = \int \frac{\gamma}{\sqrt{(1-k^2\gamma^2)^{3}}}\hspace{0.1cm} d\gamma d\theta
\end{equation}

and expand it as a series in $k^2$. So we have a expression where $F[0]$ equals the Euclidean area and $F[1]$ equals the hyperbolic area. The series expansion gives

{\scriptsize \begin{align}
F[k] = & \frac{1}{k^2} \left\{\right.-\tan ^{-1}\left(\frac{a \sin (\text{$\theta $2})}{b-a \cos (\text{$\theta $2})}\right)-\tan ^{-1}\left(\frac{b \sin (\text{$\theta $2})}{a-b \cos (\text{$\theta $2})}\right)-\cot ^{-1}\left(\frac{a \sin (\text{$\theta $2})}{b-a \cos (\text{$\theta $2})}\right)-\cot ^{-1}\left(\frac{b \sin (\text{$\theta $2})}{a-b \cos (\text{$\theta $2})}\right) \nonumber \\ 
& +\frac{1}{2} a b k^2 \sin (\text{$\theta $2}) +\frac{1}{8} a b k^4 \sin (\text{$\theta $2}) \left(a^2+a b \cos (\text{$\theta $2})+b^2\right)  \nonumber \\
& + \frac{1}{48} a b k^6 \sin (\text{$\theta $2}) \left(3 a^4+a^2 b^2 \cos (2 \text{$\theta $2})+3 a b \left(a^2+b^2\right) \cos (\text{$\theta $2})+2 a^2 b^2+3 b^4\right) \nonumber
\end{align}  }

\begin{flalign}\scriptstyle
\hspace{1.5cm} + \cdots (\text{terms proportional to the Euclidean area}) \left.\right\} &&
\end{flalign}

The first four terms in the expansion add up to zero. From the fifth term onwards, we can see that the terms are proportional to the Euclidean area $`a b sin(\theta)$', with the fifth term being equal to the Euclidean area. This means we can express the hyperbolic area as a series of Euclidean triangles.  

\subsection{Bubble}

For the Bubble, we can use eq.(\ref{seriesgengamma}) to find out the resulting expression in terms of the Euclidean area. The first term in the expansion suggests that it is proportional to the Euclidean area since $A[u_i]$ is not dependent upon $\theta$. It is given by

\begin{equation}
    {vol}_{Bubble}(\Sigma(v_i'^{\rm M})) =
    -\frac{i a b sin(\theta2) }{2(A[u_i])^{3/2}}+\cdots
\end{equation}

Now we can see that all the terms in eq.(\ref{seriesgengamma}) are proportional to $i/(A[u_i])^{3/2}$. So for the stacking up of the Feynman diagrams for the scattering amplitude happens (i.e., the triangles represented by those diagrams have common sides beginning at first order), we will multiply eq.(\ref{seriesgengamma})) by $i(A[u_i])^{3/2}$, then the result will be proportional to the original result. This proportionality can be dealt with if we shift the mass for the bubble such that

\begin{equation}
    F_{Bubble}(m_i')=i/(A[u_i])^{3/2}F_{Bubble}(m_i)
\end{equation}

where $F_{Bubble}$ is the expression for the Feynman integral of the Bubble diagram. Also we only change the mass of the propagator which is unique to only one Bubble diagram in the scattering process $1 \to 2$. Hence after doing we get the first order term equal to the Euclidean area

\begin{equation}
    {vol}_{Bubble}(\Sigma(v_i'^{\rm M})) =
    \frac{ a b sin(\theta2) }{2}+\cdots
\end{equation}

\subsection{Tadpole}

Now for the tadpole, we use the same method as for the bubble diagram. We shift the mass of the propagator which is unique to the tadpole diagram in the scattering process $1\to 2$. Doing this once again we get the first order term equal to the Euclidean area in the expansion of the resulting expression  for the tadpole diagram.

\begin{equation}
    {vol}_{Tadpole}(\Sigma(v_i'^{\rm M})) =
    \frac{ a b sin(\theta2) }{2}+\cdots
\end{equation}

\section{Stacking up}

Now we see the first order terms in the expansion of the resulting
expression of the Feynman integrals has a nice geometrical viewpoint. The triangles represented by these terms have a common side as they are Euclidean areas of the resulting triangles made by the momenta and hence can be stacked up. Note that we are not following any thorough mathematical formalism for stacking up. The whole formalism is just to show that we have common sides for the triangles using this approach. A more rigorous mathematical formalism to find out a geometry is a subject of future investigation. All these triangles are shown in Fig.(\ref{fig2}) and Fig.(\ref{fig2shift}). If we just stack up the triangles in Fig.(\ref{fig2}) using the common sides, we get the following figure, see Fig.(\ref{sup1}).

\begin{figure}[ht]
\hspace{3cm}\epsfig{file=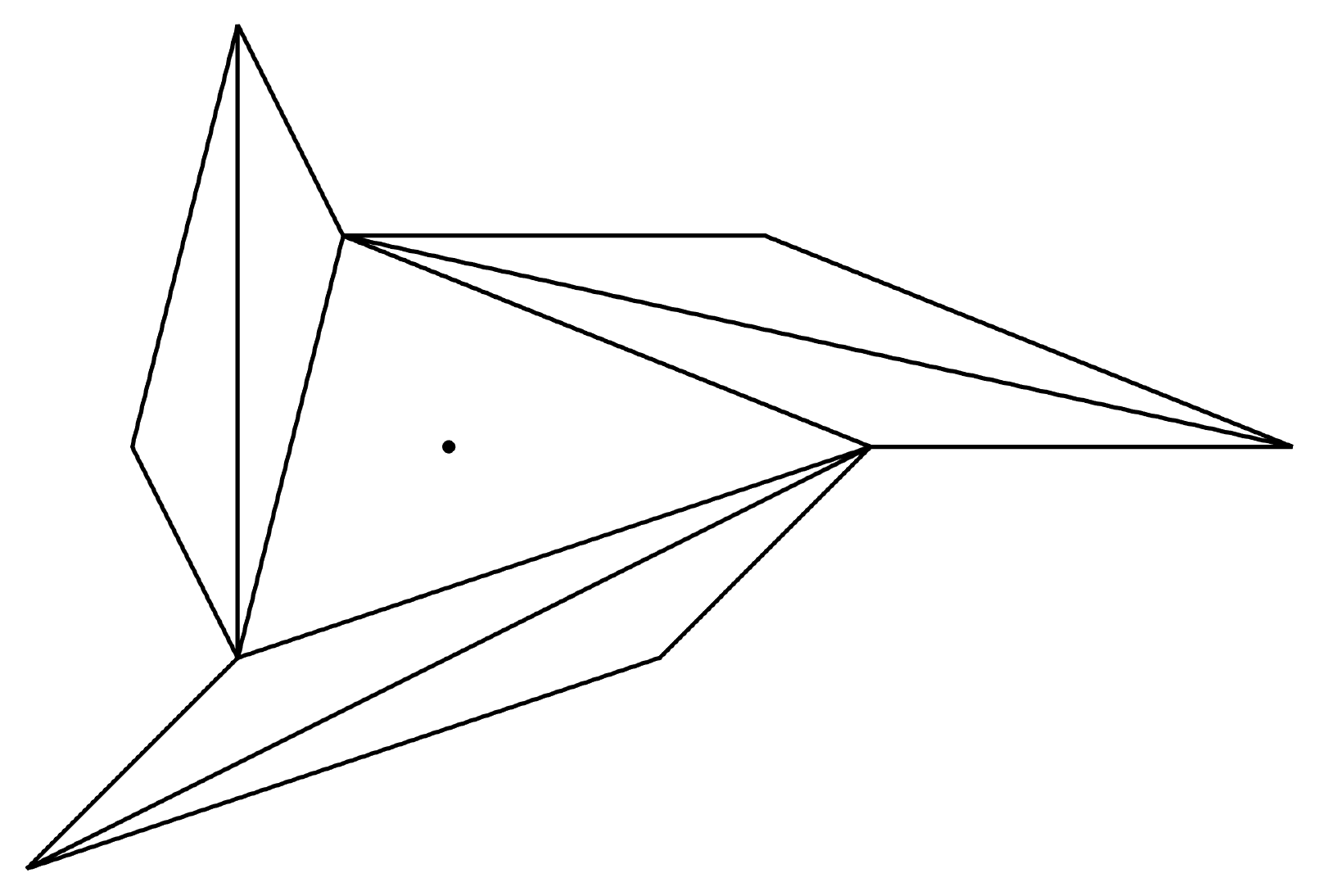,width=10cm}
\caption{Stacking up triangles - 1st step}\label{sup1}
\end{figure}

\begin{figure}[ht]
\epsfig{file=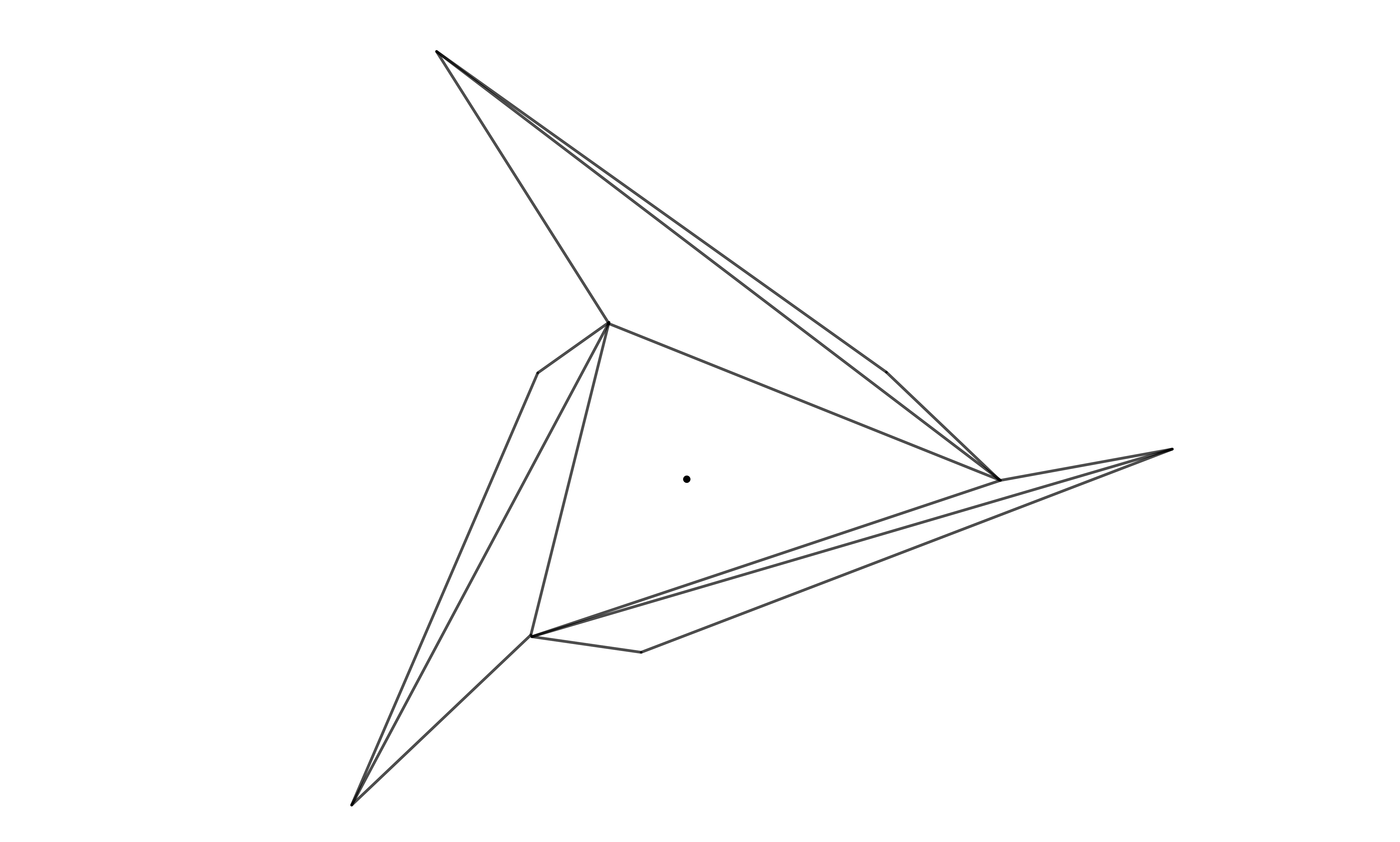,width=15cm}
\caption{Stacking up shifted triangles - 1st step}\label{sup2}
\end{figure}

Now there is one more way to do the stack up, the sides where the orientation of the `blades' of the `whirlpool wheel' type diagram can be anticlockwise rather than the clockwise orientation here. But here, we will stick to the clockwise orientation as a convention. Now if we stack up the triangles in Fig.(\ref{fig2shift}) according to the common sides approach we get, see Fig.(\ref{sup2}). So we see that the main triangle is unaffected, but the triangles corresponding to the tadpoles and the bubbles no longer have the same area but instead stack up according to their common sides and make a quadrilateral instead of a parallelogram.

But the triangles represent only the ${vol}_{C^{n-1}}(\Sigma(v_i'^{\rm M}))$ of the Feynman integral result eq.(\ref{maineq_1}), and we have a factor ${vol}_{\R^{n-1}}(\Sigma(p_i''^{\rm M}))$ in the denominator which varies with different Feynman integrals. This factor equals the Euclidean area of the corresponding Feynman integral. So actually, it's the ratio of areas of the shapes which is just a mere number. In order to see something interesting, we need to keep the area of the shape in the numerator intact. For this, we need to divide and multiply all the Feynman integrals with the area of the triangle integral. We keep the dividing triangle area in the denominator as common, and now we can add up the diagrams with a scalar multiple each and all having common denominators.

\begin{align}\label{scatamp}
\text{One-loop 3-point scattering amplitude}= \sum_{I'}\int{\rm d}^np\prod_{i\in I'}\frac{1}{Q_i'}\nonumber \\
=\frac{\hbox{vol}(S^n_{1/2})}{R\,\hbox{vol}_{\R^{n-1}}(\Sigma(p_i^{\rm M}))}\sum_{I'}\left(\frac{\hbox{vol}_{\R^{n-1}}(\Sigma(p_i^{\rm M}))}{\hbox{vol}_{\R^{n-1}}(\Sigma(p_i''^{\rm M}))}\right)\hbox{vol}_{C^{n-1}}(\Sigma(v_i'^{\rm M})).
\end{align}

So we see that the scalar multiples are the ratios of triangle areas and the areas represented by the corresponding Feynman integral. These ratios can be evaluated by first considering all the participating vectors in particular in terms of any two basis vectors out of $p_1,p_2$ and $p_3$.

\subsection{Evaluation of the ratio for the bubble and tadpole diagram}

We will first evaluate the ratios for the bubble diagram and the tad[pole diagrams. The area of the mapped triangle corresponding to the bubble diagram is given by 

\begin{equation}
\text{Area of mapped triangle for the bubble}=\frac{\bar{p}_{i,i+1}\otimes(\bar{r}_B-\bar{p}_i)}{2}
\end{equation}

and for the tadpole

\begin{equation}
\text{Area of mapped triangle for the tadpole}=\frac{(\bar{p}_i-\bar{p}_{i+2N})\otimes(\bar{p}_{i,i+1}+\bar{r}_B-\bar{p}_{i+2N})}{2}
\end{equation}

Here we can see that if we choose $\bar{p}_i\parallel\bar{p}_{i+2N}$ then the above area formula becomes most compact expression possible. Now if we rewrite $r_B$ in terms of the existing two basis vectors in the area expression, i.e.,

\begin{equation}
    \bar{r}_B=c_i^{i,i+1}\bar{p}_i+c_{i+1}^{i,i+1}\bar{p}_{i+1}
\end{equation}

then the area becomes

\begin{equation}
\text{Area of mapped triangle for the bubble}=\frac{\left(1-c_i^{i,i+1}-c_{i+1}^{i,i+1}\right)(\bar{p}_i\otimes \bar{p}_{i+1})}{2}
\end{equation}

and using the parallelity convention if

\begin{equation}
\bar{p}_i=d_{i}\hspace{0.07cm}  \bar{p}_{i+2N}
\end{equation}

then

\begin{equation}
\text{Area of mapped triangle for the tadpole}=\frac{\left(1+c_{i+1}^{i,i+1}\right)(1-d_{i})(\bar{p}_i\otimes \bar{p}_{i+1})}{2}
\end{equation}

We see that there is a common factor $(\bar{p}_i\otimes \bar{p}_{i+1})$ in both the area expressions. This is because of the fact that the  diagonals of a parallelogram divide it into four equal half areas. For example consider the parallelogram formed by $p_1$, $p_2$, $p_{12}$ and the origin

\begin{align}
    \includegraphics[height=5cm,valign=c]{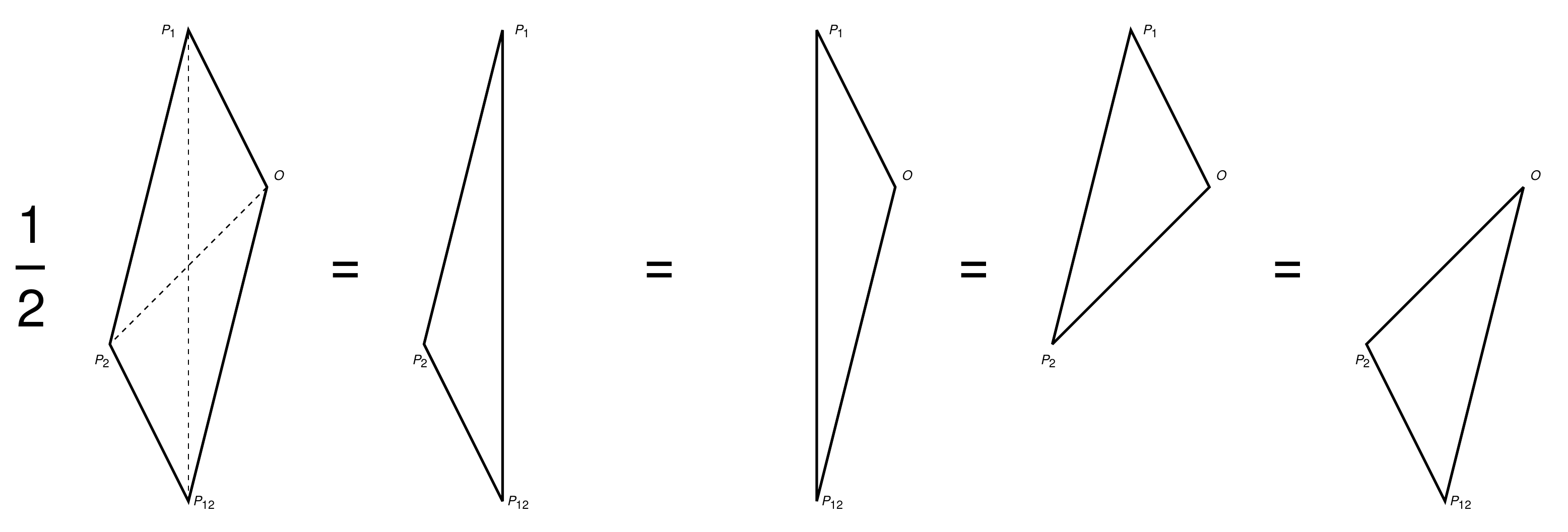}
\end{align}

Here the triangles $p_1 p_2 O, O p_2 p_{1 2}, p_1 p_2 p_{1 2} \text{and} O p_1 p_{1 2}$ all have equal areas

So there is a common factor in the scalar multiple for the bubble and the Tadpole diagrams.  For example if we consider the area factor $p_1 O p_{1 2}$ coming from the tadpole then the dividing area is equal to the area of triangle $p_1 p_2 O$ since they have equal areas. Hence the scalar multiple in this case becomes

\begin{equation}
    \frac{\hbox{vol}_{\R^{n-1}}(\Sigma(p_i^{\rm M}))}{\hbox{vol}_{\R^{n-1}}(\Sigma(p_i''^{\rm M}))}=  \frac{\text{Area of} \triangle p_1 p_2 p_3}{\left(1+c_{i+1}^{i,i+1}\right)(1-d_{i})\text{Area of} \triangle p_1 p_2 O}
\end{equation}

Now for evaluating the ratio we need to express $p_3$ in terms of $p_1$ and $p_2$. Now since they are in a plane and $p_1\neq k p_2$ we can rewrite $p_3$ as a linear combination of $p_1$ and $p_2$

\begin{equation}\label{lincom}
    p_3 = \alpha p_1 + \beta p_2
\end{equation}

then we can rewrite the scalar multiple as 

\begin{align}
    & \frac{\text{Area of} \triangle p_1 p_2 p_3}{\left(1+c_{i+1}^{i,i+1}\right)(1-d_{i})\text{Area of} \triangle p_1 p_2 O} \nonumber \\ &
    =\frac{1}{\left(1+c_{i+1}^{i,i+1}\right)(1-d_{i})}\includegraphics[height=10cm,valign=c]{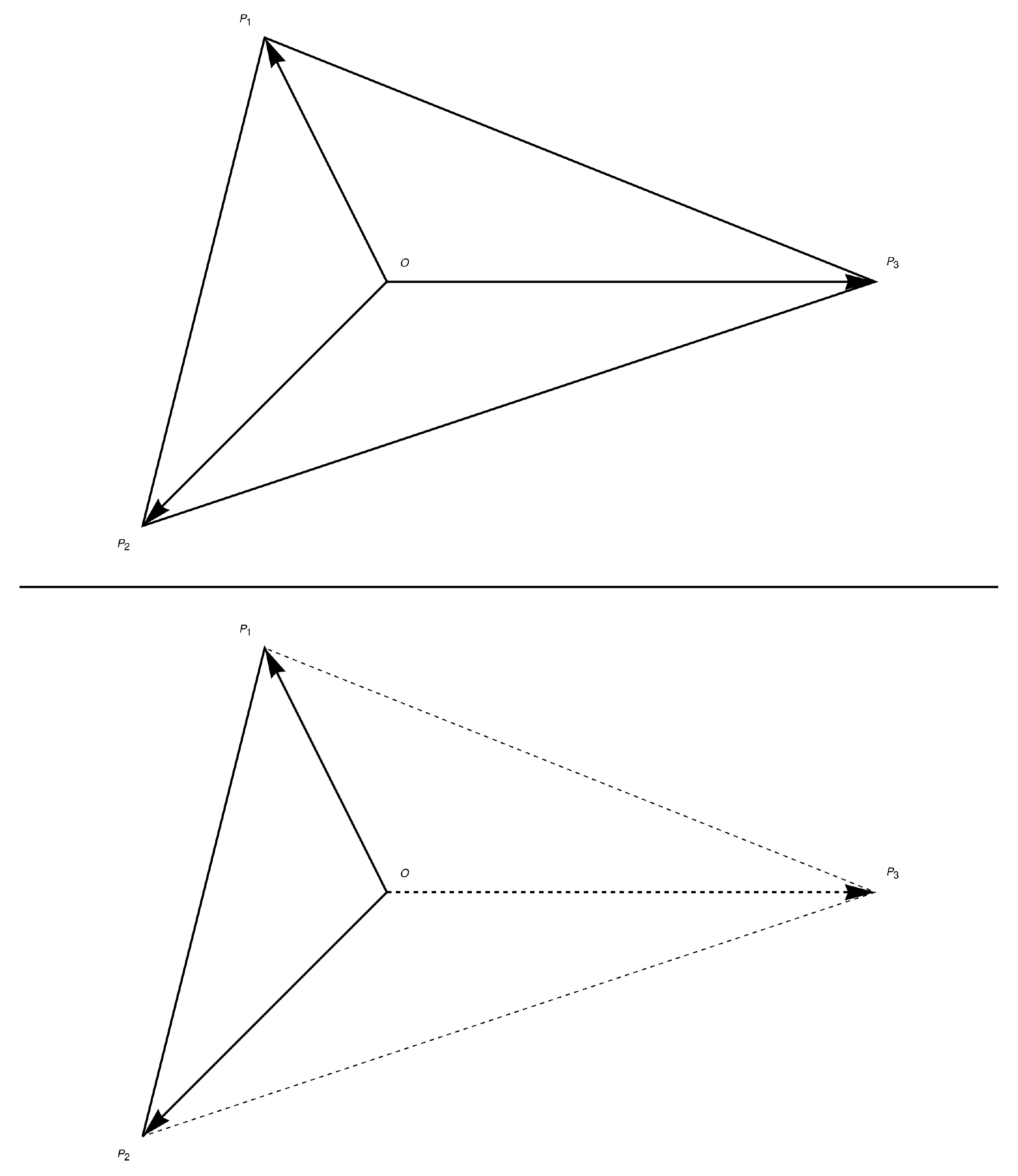} 
    \nonumber \\
      & =\frac{\text{Area of} \triangle p_1 p_2 O + \text{Area of} \triangle p_1 O p_3 + \text{Area of} \triangle O p_2 p_3 }{\left(1+c_{i+1}^{i,i+1}\right)(1-d_{i})\text{Area of} \triangle p_1 p_2 O} \nonumber
      \\
      & =\frac{p_1\otimes p_2 + p_2\otimes p_3 + p_1\otimes p_3}{\left(1+c_{i+1}^{i,i+1}\right)(1-d_{i})(p_1\otimes p_2)}
      \nonumber \\
      & = \frac{\alpha+\beta+1}{\left(1+c_{i+1}^{i,i+1}\right)(1-d_{i})} 
\end{align}

where we have used eq.(\ref{lincom}) in the last step. Now $\alpha$ and beta $\beta$ can be derived geometrically if we extend or contract $p_1$ and $p_2$ in their respective directions and try to form a triangle with $p_3$ 

\begin{figure}[ht]
\hspace{3cm}\epsfig{file=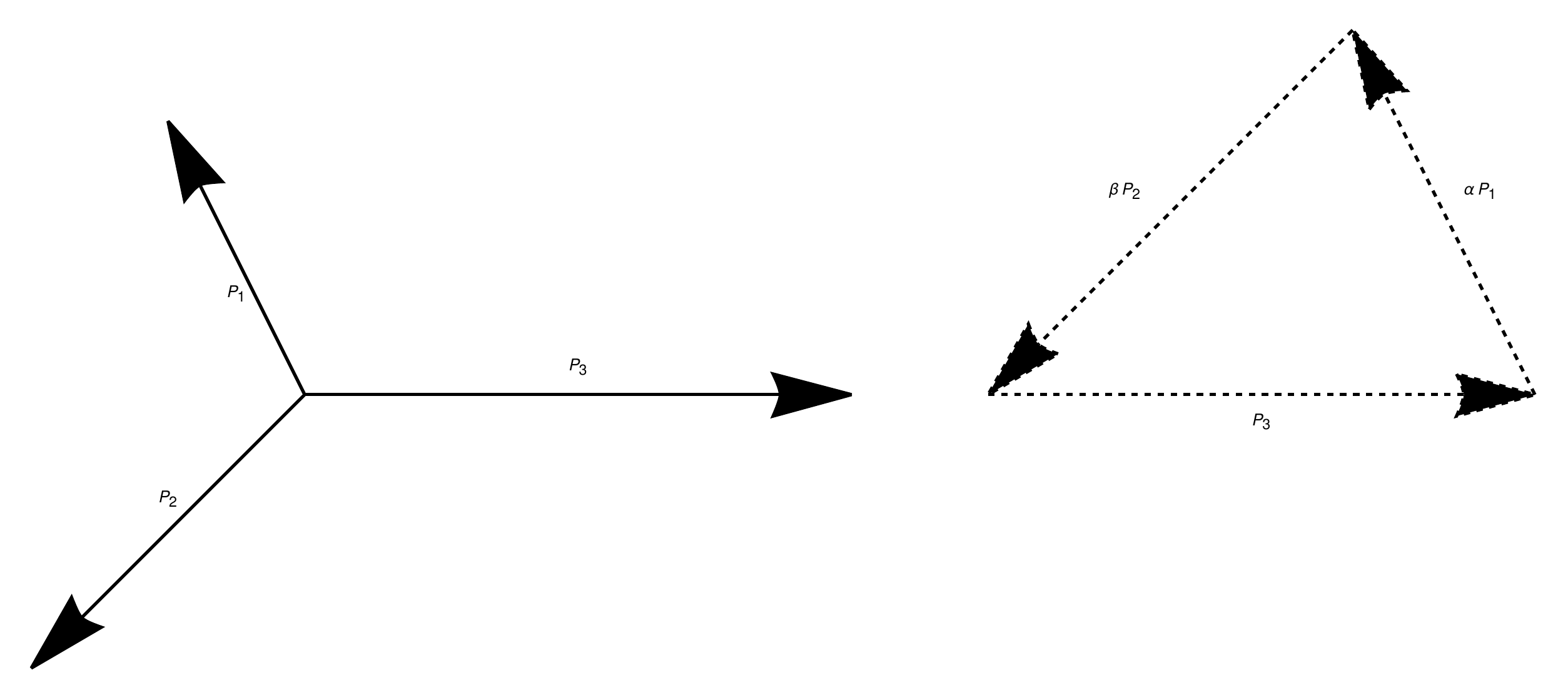,width=12.5cm}
\caption{$p_3$ as a linear combination of $p_1$ and $p_2$}\label{arrow}
\end{figure}

Now the sides of the triangle formed will correspond to the sides of the new parallelograms that needs to be added to the Fig.(\ref{sup1}).

For the tadpole case it is

 \begin{align}
    \frac{(\alpha+\beta+1)}{\left(1+c_{i+1}^{i,i+1}\right)(1-d_{i})} & (\text{Area for tadpole diagram} (\left(1+c_{i+1}^{i,i+1}\right)(1-d_{i})A(\triangle p_1 O p_{12})))\nonumber \\   & =
     (\alpha+\beta+1)( (A(\triangle p_1 O p_{12})))\nonumber \\
     & = (\alpha+\beta+1)( (A(\triangle p_1 p_2 p_{12}))) \nonumber \\
    & =(A(\triangle (\alpha p_1) p_2 p_{12}))+(A(\triangle p_1 (\beta p_2) p_{12}))+(A(\triangle p_1 p_2 p_{12}))
\end{align}

 and similarly for the bubble

\begin{align}
    \frac{(\alpha+\beta+1)}{\left(1-c_i^{i,i+1}-c_{i+1}^{i,i+1}\right)}(\text{Area for bubble diagram} (\left(1-c_i^{i,i+1}-c_{i+1}^{i,i+1}\right)A(\triangle p_1 p_2 p_{12})))\nonumber \\ = (A(\triangle (\alpha p_1) p_2 p_{12}))+(A(\triangle p_1 (\beta p_2) p_{12}))+(A(\triangle p_1 p_2 p_{12}))
\end{align}

Summing both of them gives

\begin{align}
   &  (\alpha+\beta+1)( (A(\triangle p_1 p_2 p_{12}))+ (A(\triangle p_1 O p_{12})))\nonumber \\ 
    & = (\alpha+\beta+1)(\text{Area for quadrilateral} (A(\square p_1 p_2 O p_{12}))) \nonumber \\
     & = ( (A(\square (\alpha p_1) \otimes p_{12})))+(A(\square (\beta p_2) \otimes p_{12}))+(A(\square p_1 p_2 O p_{12})) \nonumber \\
\end{align}

Diagrammatically:

\begin{align}
    \includegraphics[height=6cm,valign=c]{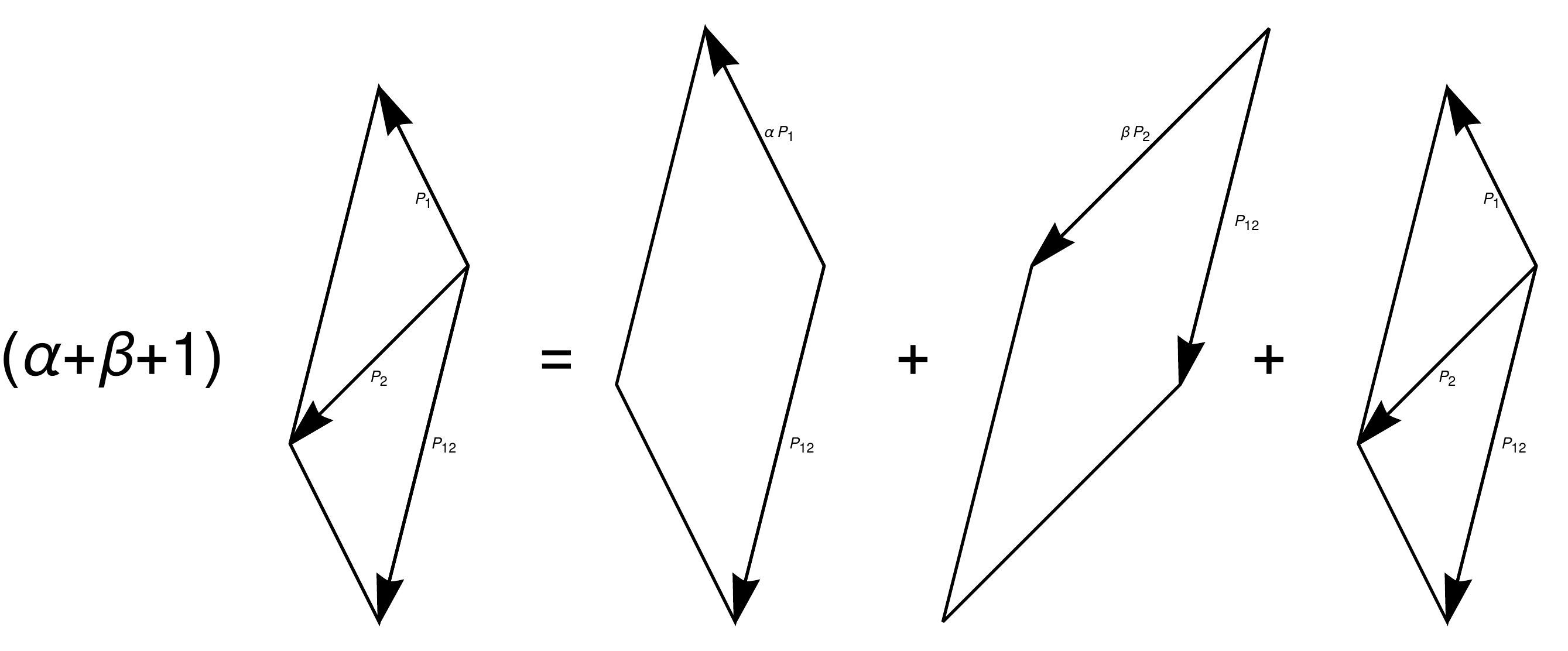} 
\end{align}
 
\begin{figure}[ht]
\hspace{3cm}\epsfig{file=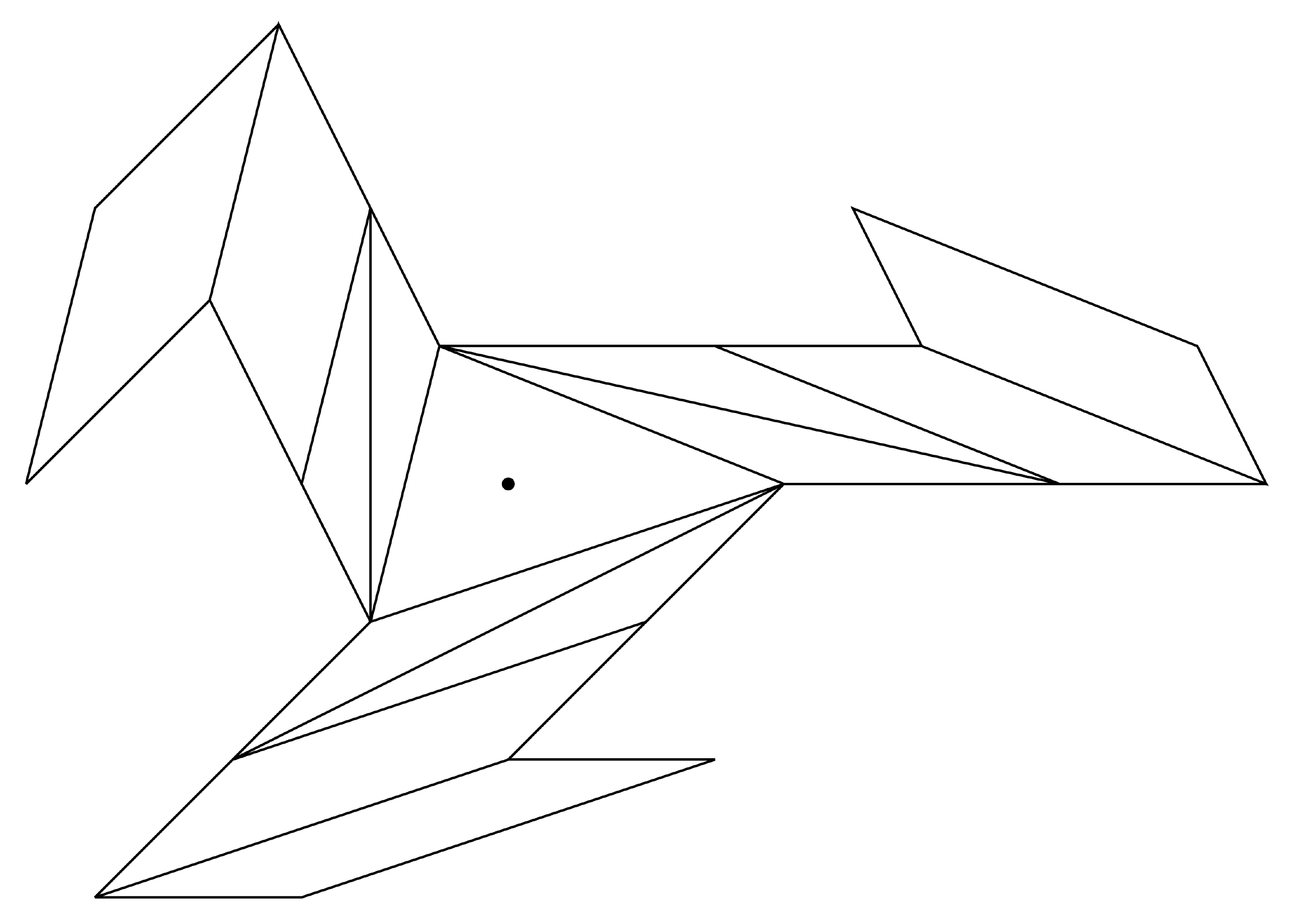,width=12.5cm}
\caption{Final figure}\label{finalsummedupfigure}
\end{figure}

So finally doing this for all the seven diagrams we finally get a figure, see Fig.(\ref{finalsummedupfigure})

Note that this is not a unique way representing the whole scattering amplitude. A more rigorous approach towards finding the precise geometry is a subject of future investigation.

	\section{Discussion and Conclusions}

 Nima Arkani Hamed and co-workers were inspired by $N=4$ SYM were able to observe that positive geometries make a natural appearance in the study of the trees and the integrands of these theories.
 Having started, they were able to consider such constructions for scattering amplitudes for bi-adjoint $\phi^3$ theory also at the tree-level.
 This gave rise to the amplituhedron (associahedron) for this theory.
 This was then extended to the construction of a halohedron wherein the integrands at one-loop were constructed. These, in turn, lead to cluster polytopes as constructed by Nima.
 The natural question then arises as to whether the one-loop integrals themselves will permit any kind of polytope construction.
 In the present work, we give an explicit construction of such a polytope construction based on the one-loop Feynman integral computation and methods introduced by Schnetz following the seminal work of Davydychev and Delbourgo \cite{Davydychev and Delbourgo(1998)}.
 In contrast to the work of associahedron and halohedron, the present work relies not on positive geometries but hyperbolic geometries.
 This feature was already observed by Schnetz.
 The hyperbolic geometry rests on eq.(57) and eq.(63) of \cite{Schnetz(2010)}.
 The independent variables and the constraints on them in the Feynman representation naturally led to their interpretations as lying in hyperbolic geometry.
 Note that the original construction does not suffice to accommodate the extension to scattering amplitudes.
 In order to achieve this end, we have had to introduce a new grammar, wherein a distinction has to be made between internal propagators carrying loop momenta and those that do not.
 In this extended system, the hyperbolic geometry becomes apparent in eq.(\ref{conic}).
 It may be noted that there are considerable algebraic simplifications that allow this to occur.
 That done, it is not yet obvious that the resulting triangles would actually produce a polytope.
 In the example at hand, this is found to be the case by explicit construction.
 Our work is motivated by the ideas using which the amplituhedron and halohedrons, and cluster polytopes are constructed. Primarily we rely on the principle of common sides of polytopes observed in Fig.(16) of \cite{Arkani-Hamed 2018}. Also, we restrict the analysis to first order series expansion of the one-loop amplitude.

 This is an introductory level work to find out a hidden geometry of scattering amplitudes at the integral level and hence yet to attain a thorough mathematical rigor and perfection. We believe that the work in section 4. and section 5. can be formulated in a more concrete mathematical framework and is a subject of future investigation.   The key achievements of our work include raising internal and loop propagators to a level where the distinction is only in terms of specific indices, eq.(\ref{inloop}), eq.(\ref{Delta'}), and eq.(\ref{finalshift}), finding a single equation which relates the masses of both internal and the loop propagators to radii of spheres, eq.(\ref{defineconstantr}), encoding the information about all the masses and the momenta and thus the information about all the parameters of the scattering amplitude in a single 2D geometrical picture which is a very fundamental and exciting way of looking at it. And finding out a general quadric surface equation, eq.(\ref{conic}), which unifies the hyperbolic geometries of the triangle, bubble, and tadpole in a single equation which is a solid proof towards a more concrete mathematical framework for the unification.
 It is, of course, natural to ask whether this can be extended to even higher orders in series expansion, loops, and other theories and subsequently becomes a work for the future.

Also, an important motivation for this work is related to the fact that the scattering amplitude is a physical observable, which is fundamentally different than the theoretical lagrangian of a quantum field theory. So this study is motivated towards finding out a geometry of the scattering amplitude itself rather than focusing on the lagrangian. The scattering amplitude, in many terms, is actually more fundamental than the lagrangian. This is supported by the studies conducted on the S-matrix bootstrap program and the recent work on positive geometries by Nima Arkani Hamed. Also, an enormous calculational simplification happens when we move from Feynman diagrams towards scattering amplitudes evident in the gluon-scattering amplitudes at the tree level giving us a hint towards the fact the scattering amplitudes are more fundamental. Our approach towards finding out a geometry at the integral level will help us go beyond the clutches of perturbation theory as we can add up the integrals at different loop orders in perturbation theory. This is a significant advancement from the positive geometry at the integrand level, as different loop orders there cannot be directly summed up. Hence our work is an important step towards discovering the geometry of scattering amplitudes up to all orders. The geometrical picture definitely hints towards finding out the scattering amplitude exactly, independent of the perturbation theory approach, as all the parameters/arguments in the scattering amplitude can be put on geometrically as we have done in our work. The S-matrix bootstrap program definitely hints towards the existence of a scattering amplitude independent of the perturbation theory and hence is in favor of our claim.

\section*{Acknowledgements}
	 AD thanks Prof.  B. Ananthanarayan and Prof. Daniel Wyler for adding valuable discussions and  comments during the course of this work. AD also thanks MHRD, India, for providing  the required funding during the course of this work.

\appendix

 \section{Appendix}

 \subsection{Table}

\begin{table}[htb!]
\begin{bigcenter}
\begin{tabular}{|cc|c|cc|}
\hline
\multicolumn{2}{|c|}{\small{Diagrams}}                            & \multirow{2}{*}{\small{$Q_i'$ involved}} & \multicolumn{2}{c|}{\small{Momenta involved in} $v_i'$}          \\ \cline{1-2} \cline{4-5} 
\multicolumn{1}{|c|}{\small{Type}}                     & \small{Identity} &                    & \multicolumn{1}{c|} {\small{$p_i'$ (Without shift)}} & \small{$p_i''$ (With shift)} \\ \hline
\multicolumn{1}{|c|}{\multirow{3}{*}{\small{Tadpole}}} & \small{T2}       & \small{$Q_2',Q_5',Q_8'$}              & \multicolumn{1}{c|}{\small{$p_2,p_{23},0$}}      & \small{$p_2,p_{23}-r_B,p_8$}    \\ \cline{2-5} 
\multicolumn{1}{|c|}{}                         & \small{T4}       & \small{$Q_3',Q_6',Q_9'$}              & \multicolumn{1}{c|}{\small{$p_3,p_{31},0$}}      & \small{$p_3,p_{31}-r_B,p_9$}    \\ \cline{2-5} 
\multicolumn{1}{|c|}{}                         & \small{T6}       &  \small{$Q_1',Q_4',Q_7'$}                   & \multicolumn{1}{c|}{\small{$p_1,p_{12},0$}}      & \small{$p_1,p_{12}-r_B,p_7$}    \\ \hline
\multicolumn{1}{|c|}{\multirow{3}{*}{\small{Bubble}}}  & \small{T3}       &  \small{$Q_2',Q_3',Q_5'$}                   & \multicolumn{1}{c|}{\small{$p_2,p_{23},p_3$}}      &  \small{$p_2,p_{23}-r_B,p_3$}   \\ \cline{2-5} 
\multicolumn{1}{|c|}{}                         & \small{T5}       &  \small{$Q_1',Q_3',Q_6'$}                   & \multicolumn{1}{c|}{\small{$p_3,p_{31},p_1$}}      &  \small{$p_3,p_{31}-r_B,p_1$}   \\ \cline{2-5} 
\multicolumn{1}{|c|}{}                         & \small{T7}       &  \small{$Q_1',Q_2',Q_4'$}                   & \multicolumn{1}{c|}{\small{$p_2,p_{12},p_1$}}      &  \small{$p_2,p_{12}-r_B,p_1$}   \\ \hline
\multicolumn{1}{|c|}{\small{Triangle}}                 & \small{T1 }      &  \small{$Q_1',Q_2',Q_3'$}                   & \multicolumn{1}{c|}{\small{$p_2,p_{3},p_1$}}      &  \small{$p_2,p_{3},p_1$}   \\ \hline
\end{tabular}
\end{bigcenter}
\caption{\label{relate}Table showing the involved $Q_i's$ and $p_i's$ for a particular Feynman diagram.}
\end{table}

\subsection{Derivation of eq.(\ref{defineconstantr})}

In \cite{Schnetz(2010)}, we saw that the constant $r$ is defined using the equation

\begin{equation}\label{22}
(p_i-c)^2+m_i^2=r^2,\quad i=1,\dots,n+1.
\end{equation}

\begin{figure}[ht]
\hspace{3cm}
\includegraphics[keepaspectratio=true, height=6cm]{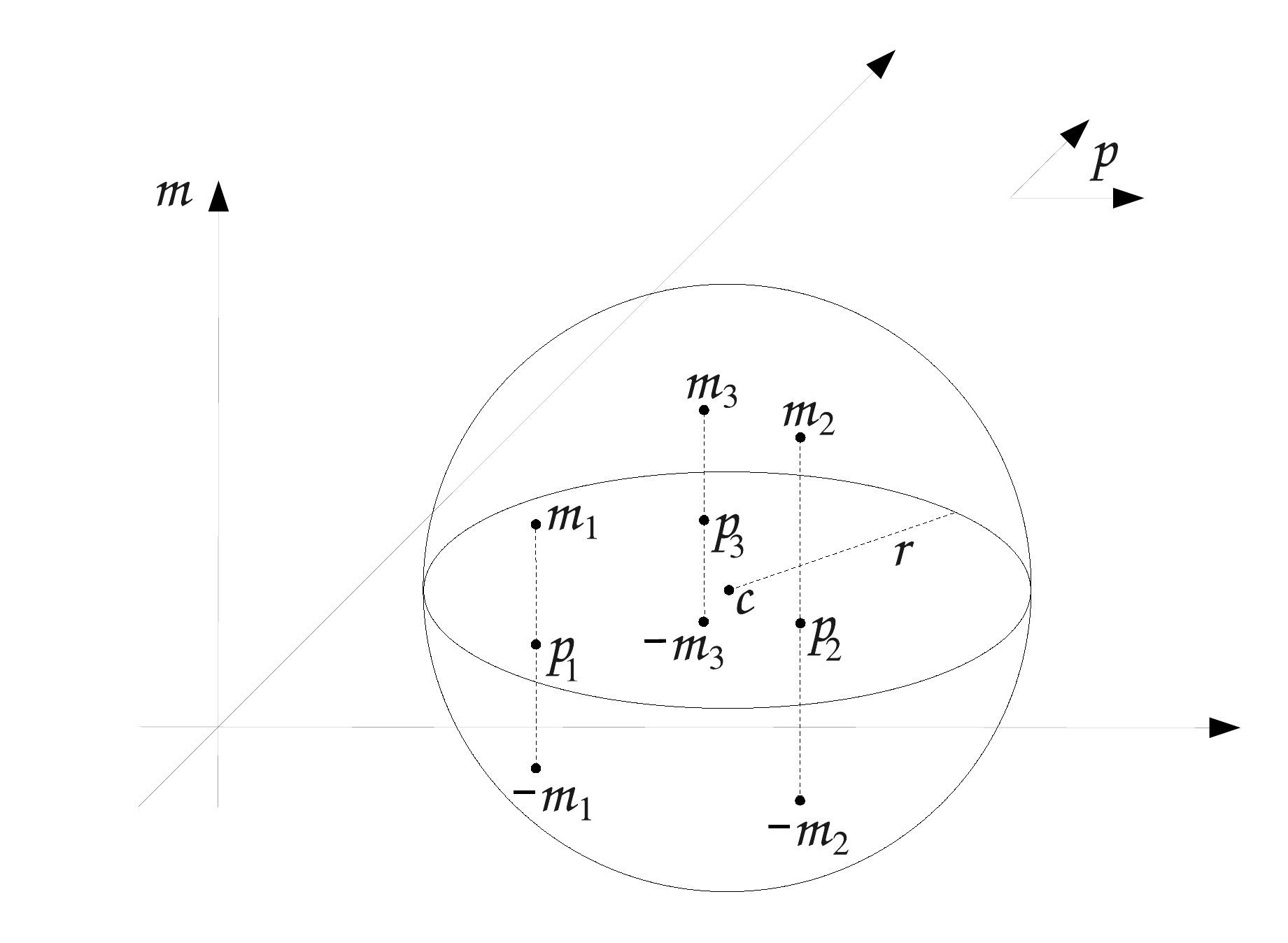}
\caption{The sphere spanned by the momenta and masses.}\label{sphere}
\end{figure}

This equation is valid since the choice of vectors $p_i$ is arbitrary such that $p_{ij}$ is fixed. Also there always exists a solution to $r$, for e.g. for the $N=3$ case, there always exists a point c on the p plane which is equidistant from the tip of the vectors $m_i$ Fig.(\ref{sphere}). 

Since $p_i$ is arbitrary we can choose point c to be the origin in our case. So eq.(\ref{22}) takes the form

\begin{equation}\label{defR_1}
p_i^2+m_i^2=r^2,\quad i=1,\dots,n+1.
\end{equation}

So in the case where $a_i=1$ the above equation is valid as the arbitrariness of the $p_i$ is only vaild here. For the other cases, i.e., $a_i=0$, the $p_i'=p_{i-N,i-N+1}$ being the external momenta are fixed. Also for case where $p_i'=0$ its a constant and hence devoid of any arbitrariness. For those cases we may write out the similar equation

\begin{equation}\label{defri}
p_i'^2+m_i'^2=r_i^2,\quad i=1,\dots,n+1.
\end{equation}

Here we see that for each internal propagator there is a corresponding constant $r_i$ and hence all the conditions of the value of internal propagators being fixed are satisfied.

Now looking at the form of eq.(\ref{Delta'}), we see that $p_i^2$ and $m_i^2$ are being multiplied by the constant coefficients $a_j$. So we need to multiply $a_j$ to eq.(\ref{defri})

\begin{equation}
 a_j^2(m_i'^2+p_i'^2)= a_j^2r_i'^2
\end{equation}

Also there are terms with the $i$ and $j$ indices switched and hence we also need the following equation:

\begin{equation}
 a_i^2(m_j'^2+p_j'^2)= a_i^2r_j'^2
\end{equation}

Finally summing the two equations, we conclude that the general equation identical to eq.(\ref{22}) in our case which will replace the $m_i's$ in eq.(\ref{Delta'}) will be of the form

\begin{equation}\label{defconr_1}
     a_i^2(m_j'^2+p_j'^2)+ a_j^2(m_i'^2+p_i'^2)= a_i^2r_j^2+a_j^2r_i'^2
\end{equation}

This equation acts on eq.(\ref{Delta'}) in a similar way like eq.(\ref{22}) acts on eq.(57) of \cite{Schnetz(2010)}. But there is a limitation in eq.(\ref{defconr_1}), if we substitute it in eq.(\ref{Delta'}) we get a dot product of the form:

$$  \frac{(a_jp_i'-a_ip_j')^2+a_j^2m_i'^2+a_i^2m_j'^2}{2} \nonumber \\ $$

\begin{align}\label{shiftedr_1}
=  \left[a_jp_i'-i\left(\sqrt{\frac{a_jr_i^2+a_ir_j^2}{2}}\right)\right]\cdot \left[a_ip_j'-i\left(\sqrt{\frac{a_jr_i^2+a_ir_j^2}{2}}\right)\right]
\end{align}

We can see that inside the term analogous to the dot product of eq.(60) in \cite{Schnetz(2010)}, $\left(\sqrt{\frac{a_jr_i^2+a_ir_j^2}{2}}\right)$ is not a constant in contrast to r there as both $a_i$ and $r_i$ change for different $i$ and $j$. Hence this can't be a valid extension of the theory presented in \cite{Schnetz(2010)}.

\begin{figure}[ht]
\hspace{0cm}
\includegraphics[keepaspectratio=true, height=5cm]{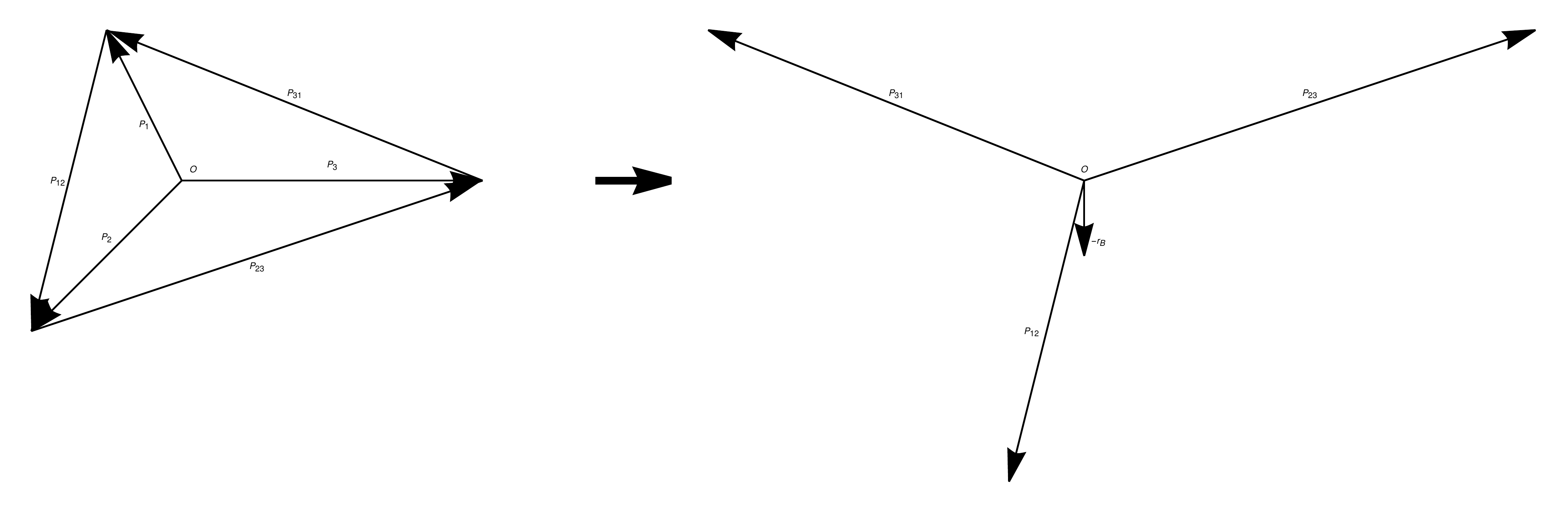}
\caption{The transformed triangle.}\label{transform}
\end{figure}

This issue can be resolved if instead of $r_i$ in the RHS of eq.(\ref{defri}) which varies with i, we have a constant say $r_{BT}$. This is possible if we try to make a similar geometrical arrangement like Fig.(\ref{sphere}) which lead to eq.(\ref{defR_1}) where the RHS is a constant. For this we once again make a triangle but now with position vectors for the vertices given by $p_{i-N,i-N+1}$ instead of $p_i$, see Fig.(\ref{transform}).

\begin{figure}[ht]
\hspace{0cm}
\includegraphics[keepaspectratio=true, height=15cm]{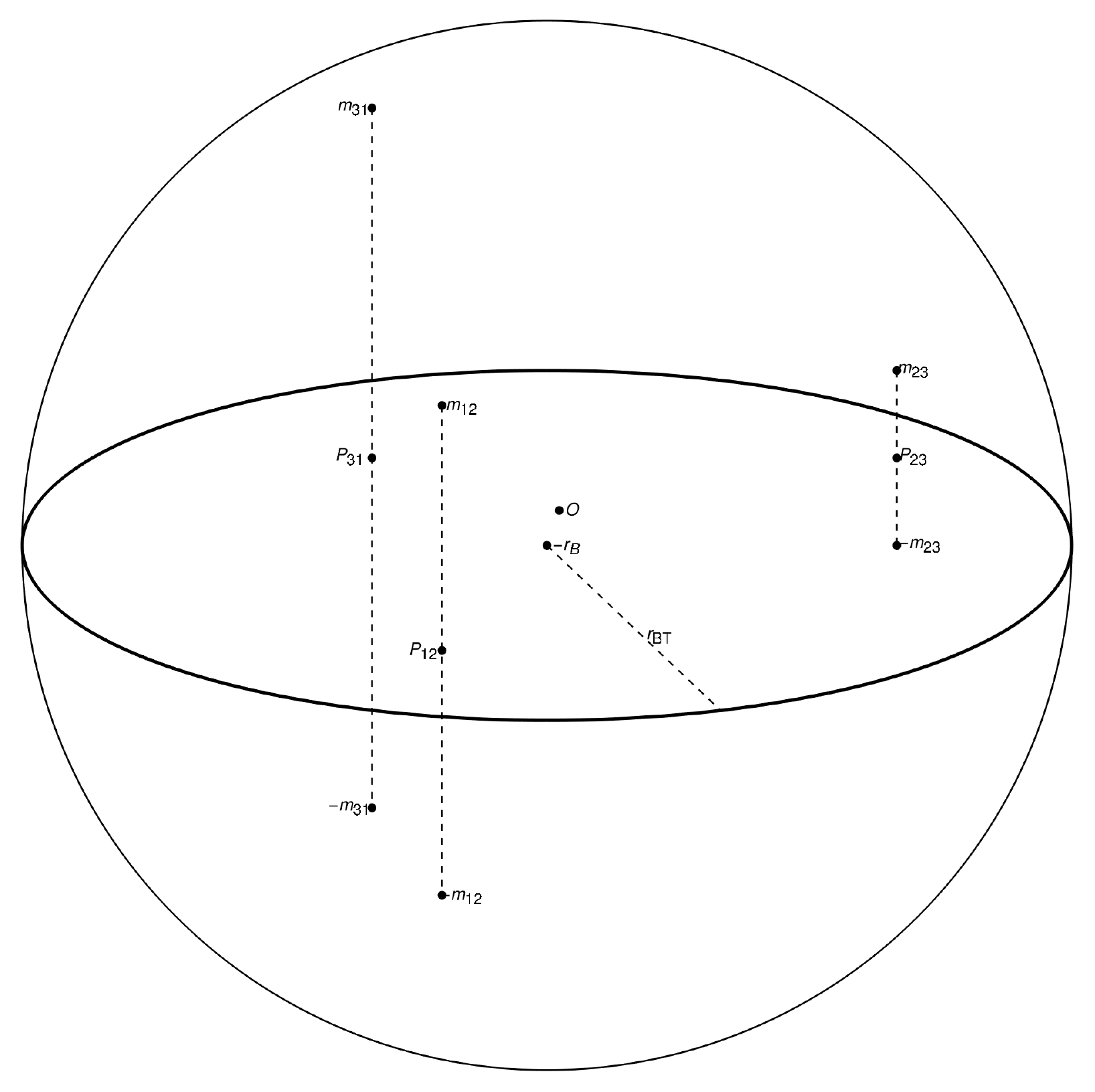}
\caption{The transformed sphere.}\label{sphere_transform}
\end{figure}

Also the corresponding $m_i$ vectors are at the vertices perpendicular to the plane of the external momenta once again making a sphere. But this time the center of the sphere is different from the origin of the position vectors of the vertices. If we denote this shift by a vector $\bar{r}_{B}$, see Fig.(\ref{sphere_transform})

then eq.(\ref{defri}) gets modified to

\begin{equation}\label{defrbt}
(\bar{p}_i'-\bar{r}_{B})^2+m_i'^2=r_{BT}^2,\quad i=1,\dots,n+1.
\end{equation}

The RHS of this equation is a constant and hence it meets the requirements of having a constant imaginary part in the terms of the dot product in eq.(\ref{shiftedr_1}). But this comes at the cost of having a shift $\bar{r}_{B}$ for the bubble case.

For the tadpole case, we can rewrite a similar equation with the RHS same as eq.(\ref{defrbt})

\begin{equation}\label{defrtad}
p_i''^2+m_i'^2=r_{BT}^2,\quad i=1,\dots,n+1.
\end{equation}

Here the value of $p_i''$ is decided by the initial condition of the value of masses $m_i$ and the constant $r_{BT}$. So if we define $p_i''$ as

\begin{equation}
   p_i''^2=
\begin{cases}
    p_i^2,& \quad   i=1,\dots ,N\\
    (p_{i-N,i-N+1}-r_{B})^2,   & \quad  i=N+1,\dots ,2N \\
    r_{BT}^2-m_i'^2 & \quad i=2N+1,\dots,3N
\end{cases}
\end{equation}

then for the $a_i=0$ case we can rewrite eq.(\ref{defconr_1}) as 

\begin{equation}
a_i^2(m_j'^2+p_j''^2)+ a_j^2(m_i'^2+p_i''^2)= (a_i^2+a_j^2)r_{BT}^2
\end{equation}

To include the $a_i=1$ case, finally we will add a term $2a_ia_jr^2$ to the RHS of the above equation:

\begin{equation}\label{defrbtr}
a_i^2(m_j'^2+p_j''^2)+ a_j^2(m_i'^2+p_i''^2)= (a_i^2+a_j^2)r_{BT}^2+2a_ia_jr^2
\end{equation}

So instead of eq.(\ref{shiftedr_1}) we have

$$ \frac{(a_jp_i''-a_ip_j'')^2+a_j^2m_i'^2+a_i^2m_j'^2}{2}  = $$

\begin{align}\label{appendshift}
 \left[a_jp_i'-ia_j\left(\sqrt{r^2+r^2_{BT}}\right)\right]\cdot \left[a_ip_j'-ia_i\left(\sqrt{r^2+r^2_{BT}}\right)\right]  + (a_i-a_j)^2r^2_{BT}
\end{align}

Now apart form the constants $a_i$, $a_j$ and an additional shift term, the RHS is exactly analogous to the dot product in \cite{Schnetz(2010)}. So this is the proper extension of the theory there.

\subsubsection{Significance of constant $r$ and the term $2a_i a_j r^2$}

Now we will try to understand the significance of the constant $r$ and the term $2a_ia_jr^2$. If we put $a_i$ and $a_j$ equal to one in eq.(\ref{defrbtr}), we get:

\begin{equation}\label{defconr_3}
     (m_j'^2+p_j''^2)+ (m_i'^2+p_i''^2)= 2r_{BT}^2+2r^2
\end{equation}

This can be thought of as being equivalent to:

\begin{equation}\label{defconr_4}
     m_i'^2+p_i''^2= r_{BT}^2+r^2
\end{equation}

Since this is valid only when all $a_i's=1$, this equation is equivalent to eq.(\ref{defR_1}) and hence $r_{BT}$ and $r$ ought to be a constant  for different values of $i$. Also for $a_i$ or $a_j=0$, eq.(\ref{defrbtr}) becomes equal to eq.(\ref{defrbt}) and eq.(\ref{defrtad}). This shows that eq.(\ref{defrbtr}) covers all the cases originating from different values of $a_i$ perfectly.

Note that the term $2a_ia_jr^2$ in  eq.(\ref{defrbtr}) satisfies different conditions originating from different values of $a_i$, the need to write the expression in eq.(\ref{Delta'}) in form of a dot product and the need to have a constant $r$ inside the complex part of the participating terms within the dot product all simultaneously. Also this term cannot be derived like the other terms in the RHS of eq.(\ref{defrbtr}). It is more of like a conjecture which satisfies all the required conditions. Hence we can say that eq.(\ref{defrbtr}) exists at a more fundamental level than eq.(\ref{22}), satisfying the more general mathematical formalism which includes the loop and internal propagators under a single framework.

\subsubsection{A hint of the existence of multidimensions in the complex plane using eq.(\ref{defrbtr})}

Looking at the form of eq.(\ref{appendshift}), we see that the constants $r$ and $r_{BT}$ appear to be perpendicular to each other within the complex plane. So eq.(\ref{defrbtr}) which is a conjecture by itself hints at the existence of multi-dimensions within the complex plane. The equation which was designed to encapsulate all the possibilities arising out of the different values of $a_i$ actually paved the way to the idea of multi-dimensions.

\subsection{Proofs of eq.(\ref{bubcon}) and eq.(\ref{tadcon})}

Now the contents of the first line of the diagonalizable matrix $A$ in eq.(\ref{diamat}) can be rewritten in a concise way as 

\begin{equation}
    A_{ij}=\alpha_k'^{(i)}a_ka_m\alpha_j'^{(m)}\left(p'^{(l)}_k\hat{l}\right)\cdot\left(p'^{(l)}_j\hat{l}\right)
\end{equation}

Similar reasoning can be given for the second line. We will discuss about it in detail later. Here $p'^{(l)}_k$ represents the $l^{th}$ component of the vector $p'_k-iR$. After the dot product evaluation this becomes

\begin{equation}
    A_{ij}=\alpha_k'^{(i)}a_ka_m\alpha_j'^{(m)}p'_{kj}
\end{equation}

where $p'_{kj}$ is the dot product of two vectors $p'_k$ and $p'_j$. Now the eigenvalue equation can be obtained by taking the determinant of this matrix subtracted by eigenvalue times the identity matrix and then equating it to zero.

\begin{equation}
    \text{Det}\left[A-\lambda I\right]=0
\end{equation}

The resulting matrix whose determinant is zero can be written in a concise way as 

\begin{equation}
    A'_{ij}=\alpha_k'^{(i)}a_ka_m\alpha_j'^{(m)}p'_{kj}-\lambda\delta_{ij}
\end{equation}

Also the detrminant of this matrix can be written in terms of the Levi-Civita tensors \cite{Riley} as 

\begin{equation}
    \text{Det}\left[ A'_{ij}\right] = \frac{1}{n!}\varepsilon_{i_1\cdots i_n}\varepsilon_{j_1\cdots j_n}A'_{i_1j_1}\cdots A'_{i_nj_n}
\end{equation}

And hence the eigenvalue equation becomes

\begin{equation}
    \varepsilon_{i_1\cdots i_n}\varepsilon_{j_1\cdots j_n}\left(\alpha_{k_1}'^{(i_1)}a_{k_1}a_{m_1}\alpha_{j_1}'^{(m_1)}p'_{k_1j_1}-\lambda\delta_{i_1j_1}\right)\cdots\left(\alpha_{k_n}'^{(i_n)}a_{k_n}a_{m_n}\alpha_{j_n}'^{(m_n)}p'_{k_nj_n}-\lambda\delta_{i_nj_n}\right)=0
\end{equation}

where summation over repeated indices is implied. Now if we consider a set

\begin{equation}
    I'=\{1',2',...,(n-s)'\} \subseteq I=\{1,2,...,n\}
\end{equation}

for $s\leq n$ and $s \in \mathbb{N}$, then the eigenvalue equation can be written as a polynomial equation in $\lambda$:

$$ \sum_{s=0}^n \lambda ^s a_{k_{1'}}a_{m_{1'}}\cdots a_{k_{(n-s)'}}a_{m_{(n-s)'}} \times $$

\begin{align}\label{polylam}
    \left(\alpha_{k_{1'}}'^{(i_{1'})}\alpha_{j_{1'}}'^{(m_{1'})}\cdots\alpha_{k_{(n-s)'}}'^{(i_{(n-s)'})}\alpha_{j_{(n-s)'}}'^{(m_{(n-s)'})}p'_{k_{1'}j_{1'}}\cdots p'_{k_{(n-s)'}j_{(n-s)'}} \varepsilon_{i_1\cdots i_n}\varepsilon_{j_1\cdots j_n}\right)=0
\end{align}

Here the bracket indicates that the summation within the bracket is considered at first.  In order to prove eq.(\ref{bubcon}), we need to show that the term with $s=0$ is zero when one of the $a_i's=0$. For this we need to show that for the $s=0$ case, only the term with full multiplicity of $a_i's$ is non-zero and all the other terms with less multiplicity cancel.

\subsubsection{Case s=0}

First of all we will consider the term with least multiplicity of $a_i's$, i.e., the term with all the $a_i's$ equal. This is possible only when all the `$a$-indices' in eq.(\ref{polylam}) are equal:

\begin{equation}
    k_{1'}=m_{1'}=k_{2'}=m_{2'}=\cdots k_{(n-s)'}=m_{(n-s)'}=\omega \hspace{1cm}  \text{...least multiplicity}  
\end{equation}

where $\omega$ is a constant and $1\leq \omega \leq n$ and the corresponding $a$, we denote by a constant $a_\omega$. So the $s=0$ term in eq.(\ref{polylam}) takes the form:

\begin{equation}\label{leastmulszero}
    a_{\omega}^{2n} \left(\alpha_{\omega}'^{(i_{1})}\alpha_{j_{1}}'^{(\omega)}\alpha_{\omega}'^{(i_{2})}\alpha_{j_{2}}'^{(\omega)}\cdots\alpha_{\omega}'^{(i_{n})}\alpha_{j_{n}}'^{(\omega)}p'_{\omega j_{1}}\cdots p'_{\omega j_{n}} \varepsilon_{i_1\cdots i_n}\varepsilon_{j_1\cdots j_n}\right)
\end{equation}

Note that for this case since we have $s=0$, the set $I'=I$ and hence we have removed the dash from all the numbers. We need to prove that the terms of this form cancel each so that the least multiplicity contribution is zero. This can be checked by using the property that the Levi-Civita tensors change their sign when we swap two of their indices. So if we swap $i_j$ by $i_k$, we can see that the expression in eq.(\ref{leastmulszero}) remains the same but the sign changes

$$a_{\omega}^{2n} \left(\alpha_{\omega}'^{(i_{1})}\alpha_{j_{1}}'^{(\omega)}\cdots\alpha_{\omega}'^{(i_{j})}\cdots\alpha_{\omega}'^{(i_{k})}\cdots\alpha_{\omega}'^{(i_{n})}\alpha_{j_{n}}'^{(\omega)}p'_{\omega j_{1}}\cdots p'_{\omega j_{n}} \varepsilon_{i_1\cdots i_j \cdots i_k \cdots i_n}\varepsilon_{j_1\cdots j_n}\right)$$

\begin{equation}
   =- a_{\omega}^{2n} \left(\alpha_{\omega}'^{(i_{1})}\alpha_{j_{1}}'^{(\omega)}\cdots\alpha_{\omega}'^{(i_{k})}\cdots\alpha_{\omega}'^{(i_{j})}\cdots\alpha_{\omega}'^{(i_{n})}\alpha_{j_{n}}'^{(\omega)}p'_{\omega j_{1}}\cdots p'_{\omega j_{n}} \varepsilon_{i_1\cdots i_k \cdots i_j \cdots i_n}\varepsilon_{j_1\cdots j_n}\right)
\end{equation}

and hence there is a counter-term which cancels every term in all possible permutations of the set ${i_1\cdots i_n}$ within  $\varepsilon_{i_1\cdots i_n}$. This is also true if we swap $j_k$ and $j_l$ from $\varepsilon_{j_1\cdots j_n}$. The corresponding terms cancel each other and hence we a zero contribution from the least multiplicity term.

Now we will consider the case where one of the indices is unequal:

\begin{equation}
    k_{1'}=m_{1'}=k_{2'}=m_{2'}=\cdots k_{(n-s)'}=\omega \neq m_{(n-s)'} = \rho \hspace{0.5cm}  \text{...second least multiplicity}
\end{equation}

For this case, the replacement of $i_j$ and $i_k$ brings the same result as the previous case for any $j$ and $k$, but the replacement of  $j_k$ and $j_l$ produces two different terms for $j/l = (n-s)'$. These different terms can be framed to resemble a single form and then cancelled only if we choose two different sets of second least multiplicity:

\begin{equation}
    k_{1'}=m_{1'}=k_{2'}=m_{2'}=\cdots k_{(n-s)'}=\omega \neq m_{k} = \rho \hspace{0.5cm} 
\end{equation}

and

\begin{equation}
    k_{1'}=m_{1'}=k_{2'}=m_{2'}=\cdots k_{(n-s)'}=\omega \neq m_{l} = \rho \hspace{0.5cm} 
\end{equation}

Then the corresponding terms cancel each other upon swapping of $j_k$ and $j_l$:

$$a_{\omega}^{2n-1}a_{\rho}  \left(\alpha_{\omega}'^{(i_{1})}\alpha_{j_{1}}'^{(\omega)}\cdots\alpha_{j_k}'^{(\rho)}\cdots\alpha_{j_{l}}'^{(\omega)}\cdots\alpha_{\omega}'^{(i_{n})}\alpha_{j_{n}}'^{(\omega)}p'_{\omega j_{1}}\cdots p'_{\omega j_{n}} \varepsilon_{i_1\cdots i_n}\varepsilon_{j_1\cdots j_k \cdots j_l \cdots  j_n}\right)$$

\begin{equation}
   =- a_{\omega}^{2n-1}a_{\rho} \left(\alpha_{\omega}'^{(i_{1})}\alpha_{j_{1}}'^{(\omega)}\cdots\alpha_{j_l}'^{(\omega)}\cdots\alpha_{j_{k}}'^{(\rho)}\cdots\alpha_{\omega}'^{(i_{n})}\alpha_{j_{n}}'^{(\omega)}p'_{\omega j_{1}}\cdots p'_{\omega j_{n}} \varepsilon_{i_1\cdots i_n}\varepsilon_{j_1\cdots j_l \cdots j_k \cdots  j_n} \right)
\end{equation}

So in this way we can cancel the second least multiplicity terms and using the similar method we can cancel the terms with higher multiplicity.

For the term term with maximum multiplicity, only specific permutations do not cancel. All other permutations can be cancelled using the above method. The permutation which does not cancel has the form for the indices:

\begin{equation}
    k_{1'}=m_{1'}=\omega_1\neq k_{2'}=m_{2'}=\omega_2\neq \cdots \neq k_{(n-s)'}=m_{(n-s)'}=\omega_{(n-s)} \hspace{0.1cm}  \text{...max multiplicity}
\end{equation}

The corresponding term in eq.(\ref{polylam}) takes the form

\begin{equation}
      a_{\omega_1}^{2}a_{\omega_2}^{2}\cdots a_{\omega_n}^{2} \left(\alpha_{\omega_1}'^{(i_{1})}\alpha_{j_{1}}'^{(\omega_1)}\alpha_{\omega_2}'^{(i_{2})}\alpha_{j_{2}}'^{(\omega_2)}\cdots\alpha_{\omega_n}'^{(i_{n})}\alpha_{j_{n}}'^{(\omega_n)}p'_{\omega_1 j_{1}}\cdots p'_{\omega_n j_{n}} \varepsilon_{i_1\cdots i_n}\varepsilon_{j_1\cdots j_n}\right)
\end{equation}

Now if we replace $i_k$ by $i_l$, then in order to get the same form first we need to replace $\omega_k$ by $\omega_l$ and then $j_k$ by $j_l$ as a second replacement. After this we not only get the same form but also with the same sign:

\begin{eqnarray}
\scriptstyle
 a_{\omega_1}^{2}\cdots a_{\omega_n}^{2} \left(\alpha_{\omega_1}'^{(i_{1})}\alpha_{j_{1}}'^{(\omega_1)}\cdots\alpha_{\omega_k}'^{(i_{k})}\alpha_{j_{k}}'^{(\omega_k)}\cdots\alpha_{\omega_l}'^{(i_{l})}\alpha_{j_{l}}'^{(\omega_l)}\cdots\alpha_{\omega_n}'^{(i_{n})}\alpha_{j_{n}}'^{(\omega_n)}p'_{\omega_1 j_{1}}\cdots p'_{\omega_n j_{n}} \varepsilon_{i_1\cdots i_k \cdots i_l \cdots i_n}\varepsilon_{j_1\cdots j_k \cdots j_l \cdots  j_n}\right) \nonumber \\
 \scriptstyle
 = a_{\omega_1}^{2}\cdots a_{\omega_n}^{2} \left(\alpha_{\omega_1}'^{(i_{1})}\alpha_{j_{1}}'^{(\omega_1)}\cdots\alpha_{\omega_l}'^{(i_{l})}\alpha_{j_{l}}'^{(\omega_l)}\cdots\alpha_{\omega_k}'^{(i_{k})}\alpha_{j_{k}}'^{(\omega_k)}\cdots\alpha_{\omega_n}'^{(i_{n})}\alpha_{j_{n}}'^{(\omega_n)}p'_{\omega_1 j_{1}}\cdots p'_{\omega_n j_{n}} \varepsilon_{i_1\cdots i_l \cdots i_k \cdots i_n}\varepsilon_{j_1\cdots j_l \cdots j_k \cdots  j_n}\right)
 \end{eqnarray}

So this means that the terms of this form don't cancel and are the only species which contribute to the $s=0$ term. These are the maximum multiplicity terms and hence if any of the $a_i's$ is zero, this term will consequently vanish. This will lead to eq.(\ref{polylam}) to be of the form:

$$ \sum_{s=1}^n \lambda ^s a_{k_{1'}}a_{m_{1'}}\cdots a_{k_{(n-s)'}}a_{m_{(n-s)'}} \times $$

\begin{align}
    \left(\alpha_{k_{1'}}'^{(i_{1'})}\alpha_{j_{1'}}'^{(m_{1'})}\cdots\alpha_{k_{(n-s)'}}'^{(i_{(n-s)'})}\alpha_{j_{(n-s)'}}'^{(m_{(n-s)'})}p'_{k_{1'}j_{1'}}\cdots p'_{k_{(n-s)'}j_{(n-s)'}} \varepsilon_{i_1\cdots i_n}\varepsilon_{j_1\cdots j_n}\right)=0
\end{align}

So we can take out a $\lambda$ term common and hence one of the solutions of this equation will always be a trivial solution. This proves eq.(\ref{bubcon}). Similar reasoning goes for the second line terms in eq.(\ref{diamat}). They also give rise to hyperboloid with one flat axis and hyperboloid with two flat axis and hence consequently producing hyperboloid with one flat axis as the final result. We need to dig deeper to higher values of s in order to generalize the proof. The next proof for eq.(\ref{tadcon}) will require the $s=1$ term.

\subsubsection{Case s=1}

Here we will first consider the term with maximum multiplicity. 

\begin{equation}\scriptstyle
    k_{1'}=m_{1'}=\omega_{1'}\neq k_{2'}=m_{2'}=\omega_{2'}\neq \cdots \neq k_{(n-s)'}=\omega_{(n-s)'}\neq m_{(n-s)'}=\omega_{(n-s+1)'} \hspace{0.1cm}  \text{...max multiplicity}
\end{equation}

Note that we have retained the dash in the indices of omega as $I'$ is a proper subset of $I$ for this case. The corresponding term in eq.(\ref{polylam}) is given by:

$$    a_{\omega_{1'}}^{2}a_{\omega_{2'}}^{2}\cdots a_{\omega_{(n-2)'}}^2 a_{\omega_{(n-1)'}}a_{\omega_{n'}}\lambda \times$$

\begin{equation}\scriptstyle
   \left(\alpha_{\omega_{1'}}'^{(i_{1'})}\alpha_{j_{1'}}'^{(\omega_{1'})}\alpha_{\omega_{2'}}'^{(i_{2'})}\alpha_{j_{2'}}'^{(\omega_{2'})}\cdots\alpha_{\omega_{(n-2)'}}'^{(i_{(n-2)'})}\alpha_{j_{(n-2)'}}'^{(\omega_{(n-2)'})}\alpha_{\omega_{(n-1)'}}'^{(i_{n'})}\alpha_{j_{n'}}'^{(\omega_{n'})}p'_{\omega_{1'} j_{1'}}\cdots p'_{\omega_{(n-1)'} j_{(n-1)'}} \varepsilon_{i_1\cdots i_n}\varepsilon_{j_1\cdots j_n}\right)
\end{equation}

Here also we can use the method in the previous case for the maximum multiplicity. We swap $i_{k'}$ by $i_{l'}$, then consecutively swapping $\omega_{k'}$ by $\omega_{l'}$ and $j_{k'}$ by $j_{l'}$, we get the same form and again with the same sign. 

$$    a_{\omega_{1'}}^{2}a_{\omega_{2'}}^{2}\cdots a_{\omega_{k'}}^{2}\cdots a_{\omega_{l'}}^{2}\cdots a_{\omega_{(n-2)'}}^2 a_{\omega_{(n-1)'}}a_{\omega_{n'}}\lambda \times$$

\begin{equation*}\scriptstyle
   \left(\alpha_{\omega_{1'}}'^{(i_{1'})}\alpha_{j_{1'}}'^{(\omega_{1'})}\cdots\alpha_{\omega_{k'}}'^{(i_{k'})}\cdots\alpha_{\omega_{l'}}'^{(i_{l'})}\cdots\alpha_{\omega_{(n-1)'}}'^{(i_{n'})}\alpha_{j_{n'}}'^{(\omega_{n'})}p'_{\omega_{1'} j_{1'}}\cdots p'_{\omega_{(n-1)'} j_{(n-1)'}}\varepsilon_{i_1\cdots i_k \cdots i_l \cdots i_n}\varepsilon_{j_1\cdots j_k \cdots j_l \cdots  j_n}\right)
\end{equation*}

$$   = a_{\omega_{1'}}^{2}a_{\omega_{2'}}^{2}\cdots a_{\omega_{l'}}^{2}\cdots a_{\omega_{k'}}^{2}\cdots a_{\omega_{(n-2)'}}^2 a_{\omega_{(n-1)'}}a_{\omega_{n'}}\lambda \times$$

\begin{equation}\scriptstyle
   \left(\alpha_{\omega_{1'}}'^{(i_{1'})}\alpha_{j_{1'}}'^{(\omega_{1'})}\cdots\alpha_{\omega_{l'}}'^{(i_{l'})}\cdots\alpha_{\omega_{k'}}'^{(i_{k'})}\cdots\alpha_{\omega_{(n-1)'}}'^{(i_{n'})}\alpha_{j_{n'}}'^{(\omega_{n'})}p'_{\omega_{1'} j_{1'}}\cdots p'_{\omega_{(n-1)'} j_{(n-1)'}}\varepsilon_{i_1\cdots i_l \cdots i_k \cdots i_n}\varepsilon_{j_1\cdots j_l \cdots j_k \cdots  j_n}\right)
\end{equation}

Hence this term has a non-zero contribution to the $s=1$ term in eq.(\ref{polylam}). Now we will consider the next to maximum multiplicity or second maximum multiplicity term. The indices for this case are given by 

\begin{equation}\scriptstyle
    k_{1'}=m_{1'}=\omega_{1'}\neq k_{2'}=m_{2'}=\omega_{2'}\neq \cdots \neq k_{(n-s)'}=m_{(n-s)'}=\omega_{(n-s)'} \hspace{0.1cm}  \text{... next to max multiplicity}
\end{equation}

and the corresponding term is given by

$$    a_{\omega_{1'}}^{2}a_{\omega_{2'}}^{2}\cdots a_{\omega_{(n-2)'}}^2 a_{\omega_{(n-1)'}}^2\lambda \times$$

\begin{equation}\scriptstyle
   \left(\alpha_{\omega_{1'}}'^{(i_{1'})}\alpha_{j_{1'}}'^{(\omega_{1'})}\alpha_{\omega_{2'}}'^{(i_{2'})}\alpha_{j_{2'}}'^{(\omega_{2'})}\cdots\alpha_{\omega_{(n-1)'}}'^{(i_{(n-1)'})}\alpha_{j_{(n-1)'}}'^{(\omega_{(n-1)'})}p'_{\omega_{1'} j_{1'}}\cdots p'_{\omega_{(n-1)'} j_{(n-1)'}} \varepsilon_{i_1\cdots i_n}\varepsilon_{j_1\cdots j_n}\right)
\end{equation}

Here also we can apply the previous method for maximum multiplicity and conclude that the contribution of this second maximum multiplicity term is non-zero for the $s=1$ term in eq.(\ref{polylam}). Now consequently we will move on to next to next maximum multiplicity term with indices:

\begin{equation}\scriptstyle
    k_{1'}=m_{1'}=\omega_{1'}\neq k_{2'}=m_{2'}=\omega_{2'}\neq \cdots \neq k_{(n-s)'}=m_{(n-s)'}=\omega_{1'} \hspace{0.1cm}  \text{... third max multiplicity}
\end{equation}

Note that here we are considering only a particular permutation term of the third maximum multiplicity. The corresponding term is given by

$$    a_{\omega_{1'}}^{4}a_{\omega_{2'}}^{2}\cdots a_{\omega_{(n-2)'}}^2\lambda \times$$

\begin{equation}\scriptstyle
   \left(\alpha_{\omega_{1'}}'^{(i_{1'})}\alpha_{j_{1'}}'^{(\omega_{1'})}\cdots\alpha_{\omega_{(n-2)'}}'^{(i_{(n-2)'})}\alpha_{j_{(n-2)'}}'^{(\omega_{(n-2)'})}\alpha_{\omega_{1'}}'^{(i_{(n-1)'})}\alpha_{j_{(n-1)'}}'^{(\omega_{1'})}p'_{\omega_{1'} j_{1'}}\cdots p'_{\omega_{(n-1)'} j_{(n-1)'}} \varepsilon_{i_1\cdots i_n}\varepsilon_{j_1\cdots j_n}\right)
\end{equation}

Now if we replace $i_{1'}$ by $i_{(n-1)'}$ then the expression devoid of the Levi-Civita tensor remains the same but with a different sign and hence they cancel each other. This means the contribution of this third maximum multiplicity term is zero. We can extend the statement to include the third maximum multiplicity terms of different permutation and infer that they all cancel each other rendering a vanishing contribution to the $s=1$ term.

Also owing to the similarity in the endoskeletal structure of the lower multiplicity terms, the argument can be easily extended to these terms and we can conclude that all of them do not have any contribution. This means only the maximum and next to maximum multiplicity terms have a contribution to the $s=1$ term and hence if any two of the $a_i's$ are zero, the $s=1$ term will vanish. And hence eq.(\ref{polylam}) will become of the form:

$$ \sum_{s=2}^n \lambda ^s a_{k_{1'}}a_{m_{1'}}\cdots a_{k_{(n-s)'}}a_{m_{(n-s)'}} \times $$

\begin{align}
    \left(\alpha_{k_{1'}}'^{(i_{1'})}\alpha_{j_{1'}}'^{(m_{1'})}\cdots\alpha_{k_{(n-s)'}}'^{(i_{(n-s)'})}\alpha_{j_{(n-s)'}}'^{(m_{(n-s)'})}p'_{k_{1'}j_{1'}}\cdots p'_{k_{(n-s)'}j_{(n-s)'}} \varepsilon_{i_1\cdots i_n}\varepsilon_{j_1\cdots j_n}\right)=0
\end{align}

Here we can take out $\lambda^2$ common and consequently two of the solutions will be trivial. This proves eq.(\ref{tadcon}). Once again the second line terms in eq.(\ref{diamat}) give rise to hyperboloid with two flat axes and hence the final result remains the same. Now for the cases of $s\geq2$, we can extend the arguments of the proof and conclude that $s$ of the solutions will be trivial.

\subsubsection{General s}

The arguments above can be easily extended to include the case of general $s$. All the higher multiplicity terms up to $s^{th}$ maximum multiplicity follow the arguments of proofs presented above. Since we don't have a common $\omega$ whenever we replace $i_{k'}'s$, we have to replace $\omega_{k'}'s$ and followed by $j_{k'}'s$ in order to get a similar form but with the same sign. And the moment we get the $(s+1)^{th}$ maximum multiplicity term, we have a common $\omega$, and the replacement of the $i_{k'}'s$ doesn't matter anymore, which leads to the cancellation of these kinds of terms. So the coefficient of the $\lambda^s$ term in eq.(\ref{polylam}) contains terms with at least $s^{th}$ maximum multiplicity and higher ones. And hence they have $s$ trivial solutions.

\subsection{External mathematica codes}

Here we give the details of the mathematica codes we have provided in order to reproduce Fig.(\ref{sup2}). So given the external momenta and the masses of all the loop and internal propagators, we can reproduce the final figure. First of all, we reproduce the triangles of Fig.(\ref{fig2}) and Fig.(\ref{fig2shift}) corresponding to the triangle, bubble, and tadpole Feynman diagrams in Triangle.nb, Bubble.nb, and Tadpole.nb, respectively. Then we add up all the contributions in Finalfig.nb. Finally, we give the corresponding quadric surface diagrams in the notebooks.

\subsubsection{Triangle.nb}

This is the first notebook to be read in the series of the notebooks. Here we have described that, given the credentials of the external momenta $p_{ij}$ and the loop propagator masses $m_i$, how to find the center `c' of the original sphere in eq.(\ref{22}) inside the triangle corresponding to the triangle Feynman diagram. In order to do that, we use the property that the center `c' is equidistant from the tip of the mass vectors in Fig.(\ref{sphere}). Then we set $c=0$ following our convention of keeping the center at the origin. Using this, we find out the coordinates of the vertices of our triangles with reference to the origin as the center of the sphere. In this way, we get the triangle corresponding to the triangle Feynman diagram.

\subsubsection{Bubble.nb}

For the bubble Feynman diagrams, we use the property that two of the vertices of the triangle corresponding to them coincide with the triangle corresponding to the triangle Feynman diagram. The third vertex is given by $p_{ij}-r_B$ if the first two vertices are given by $p_i$ and $p_j$. The shift $r_B$ is decided by the masses of the internal propagators from the bubble Feynman diagrams, see eq.(\ref{defrbt}). Finally, we take the mirror image of the third vertex about the line joining the first two vertices so that the triangle corresponding to the bubble Feynman diagram falls outside the original triangle when all the triangles are stacked up in the final figure, see Finalfig.nb.

\subsubsection{Tadpole.nb}

After the triangles corresponding to the bubble and the triangle Feynman diagrams are created (Triangle.nb, Bubble.nb), we can construct the triangles corresponding to the tadpole Feynman diagrams. One of the vertex $p_i$ is the same as the original triangle and the bubble, as evident in Table 1, one more vertex is the same as the bubble $p_{ij}-r_B$ and the third vertex vector is proportional to the first vertex vector with magnitude depending upon the internal propagator mass. We have denoted it here by F1. Finally, we take the reflection of the second and the third about the line $p_{ij}$ so that the tadpole triangle falls outside the original triangle when all the triangles are stacked up in the final figure, see Finalfig.nb.

\subsection{Finalfig.nb}

Finally, we stack up all the triangles in order to get the final figure in Fig.(\ref{sup2}). Note that we are not reproducing Fig.(\ref{finalsummedupfigure}) where the denominator consideration is taken care of.

\end{document}